\documentclass[aps,prd,twocolumn,amsmath,superscriptaddress,amssymb,showpacs,floatfix,nofootinbib,longbibliography]{revtex4-1}
\pdfoutput=1
\usepackage{graphicx}
\usepackage{bm}
\usepackage{times}
\usepackage{hyperref}
\usepackage{slashed}
\usepackage{color}
\usepackage{aas_macros}
\usepackage{slashed}
\usepackage{lipsum}
\usepackage{subfigure}
\usepackage{multirow}
\usepackage{amsmath}
\usepackage{array} 
\usepackage{varwidth} 
\usepackage[dvipsnames]{xcolor}
\usepackage[left]{lineno}
\usepackage{enumitem}

\definecolor{orange_cooper_gorfer}{HTML}{D97904}

\hypersetup{
    colorlinks   = true,
    citecolor    = orange_cooper_gorfer,
    urlcolor     = orange_cooper_gorfer,
    linkcolor    = orange_cooper_gorfer
}

\newcommand{\be}{\begin{equation}}
\newcommand{\ee}{\end{equation}}
\newcommand{\bea}{\begin{eqnarray}}
\newcommand{\eea}{\end{eqnarray}}

\let\vec\bm

\newcommand{\diff}{\ensuremath{\mathrm{d}}}
\newcommand{\e}{\mathrm{e}}

\newcommand{\ddelta}{\delta_\mathrm{D}}

\newcommand{\step}{\theta_\mathrm{H}}

\newcommand{\ac}{a_\mathrm{coll}}
\newcommand{\rc}{r_\mathrm{core}}
\newcommand{\Rc}{r_\mathrm{cusp}}

\newcommand{\rhoc}{\bar\rho_0}
\newcommand{\sigmav}{\langle\sigma v\rangle}
\newcommand{\Td}{T_\mathrm{d}}
\newcommand{\ad}{a_\mathrm{d}}
\newcommand{\kfs}{k_\mathrm{fs}}
\newcommand{\rhoeffo}{\langle\rho_\mathrm{eff}\rangle_0}

\newcommand{\mDM}{m_\mathrm{DM}}
\newcommand{\MDM}{M}
\newcommand{\gfac}{f_\mathrm{surv}}
\newcommand{\lfac}{f_\mathrm{tidal}}

\newcommand{\rp}{r_\mathrm{p}}
\newcommand{\df}{f_\mathrm{DF}}

\graphicspath{{Figures/}}

\newcommand\orcid[1]{\href{https://orcid.org/#1}{$\!$\includegraphics[scale=0.0045]{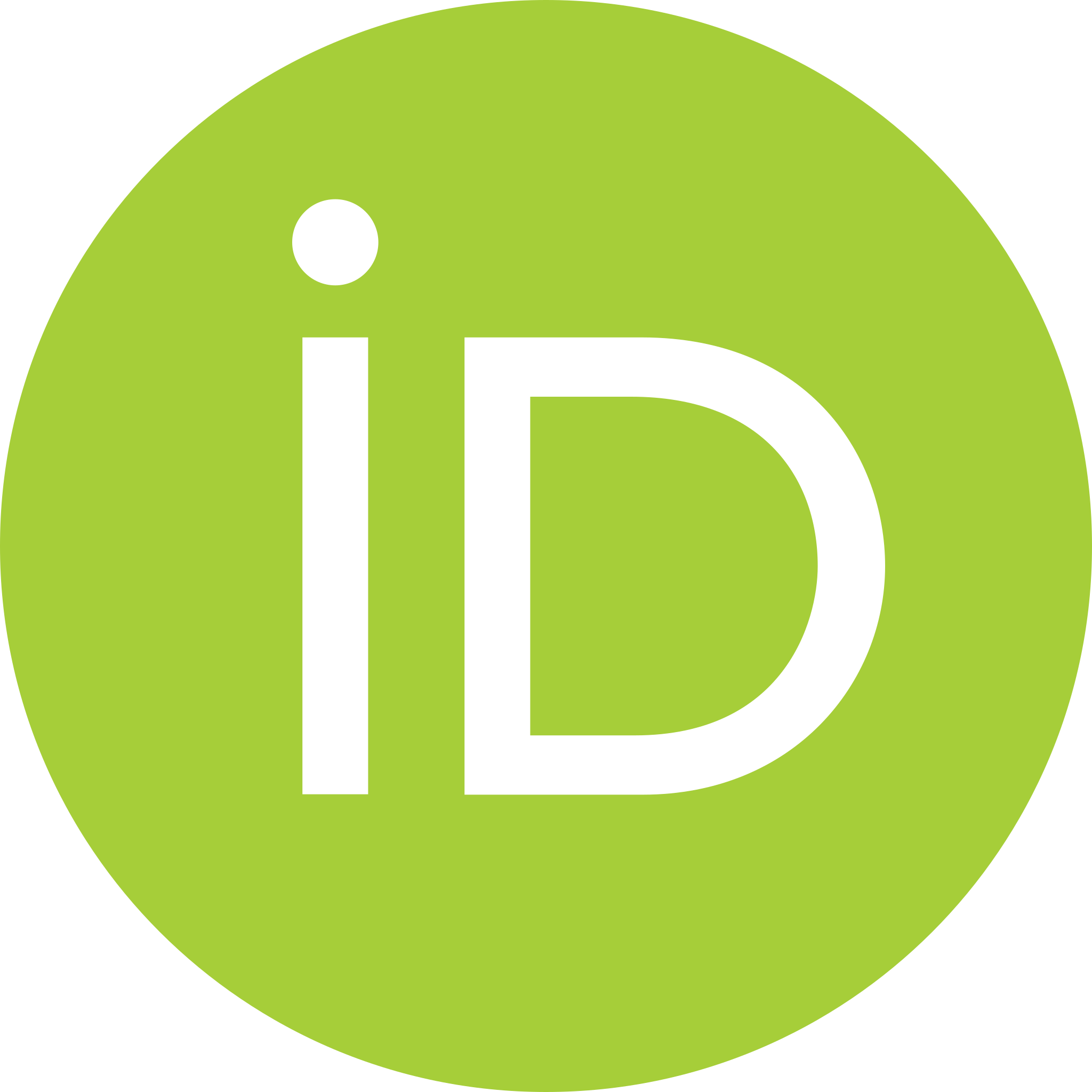} $\!\!$}}

\begin{document}

\title{Gamma-Ray Observations of Galaxy Clusters \\ Strongly Constrain Dark Matter Annihilation in Prompt Cusps}
\author{Milena Crnogor\v{c}evi\'{c} 
\orcid{0000-0002-7604-1779}}
\email{milena.crnogorcevic@fysik.su.se}
\affiliation{Stockholm University and The Oskar Klein Centre for Cosmoparticle Physics,  Alba Nova, 10691 Stockholm, Sweden}
\author{M. Sten Delos
\orcid{0000-0003-3808-5321}}
\email{mdelos@carnegiescience.edu}
\affiliation{Carnegie Observatories, 813 Santa Barbara Street, Pasadena, CA 91101, USA}
\author{Nadia Kuritz{\'e}n
\orcid{0009-0009-1670-6247}}
\email{nadia.kuritzen@gmail.com}
\affiliation{Stockholm University and The Oskar Klein Centre for Cosmoparticle Physics,  Alba Nova, 10691 Stockholm, Sweden}
\affiliation{The Viktor Rydberg Schools Foundation and Viktor Rydberg Gymnasium Odenplan, Frejgatan 30, 11349 Stockholm, Sweden}
\author{Tim Linden
\orcid{0000-0001-9888-0971}}
\email{linden@fysik.su.se}
\affiliation{Stockholm University and The Oskar Klein Centre for Cosmoparticle Physics,  Alba Nova, 10691 Stockholm, Sweden}
\affiliation{Erlangen Centre for Astroparticle Physics (ECAP), Friedrich-Alexander-Universität \\ Erlangen-Nürnberg, Nikolaus-Fiebiger-Str. 2,
91058 Erlangen, Germany}

\begin{abstract}
Thermal dark matter models generically include the prompt creation of highly-concentrated dark matter cusps in the early Universe. Recent studies find that these cusps can survive to the present day, as long as they do not fall into extremely dense regions of baryonic structure. In this work, we build models of dark matter annihilation within the prompt cusps that reside in galaxy clusters, showing that they dominate the total $\gamma$-ray annihilation signal. Using 15~years of \textit{Fermi}-LAT data, we find no evidence for a $\gamma$-ray excess from these sources, and set strong constraints on 
annihilating dark matter. These constraints generically rule out the thermal annihilation cross-section to the $b\bar{b}$ channel for dark matter masses below $\sim$200~GeV.
\end{abstract}

\maketitle

\section{Introduction}
\vspace{-0.2cm}
In standard cosmological models, ``prompt cusps'' of high dark matter density may efficiently form during certain phases in the early Universe~\cite{Diemand:2005wv,Ishiyama:2010,Anderhalden:2013wd,Ishiyama:2014uoa,Polisensky:2015,Ogiya:2016hyo,Angulo:2016qof,Delos:2017thv,Delos:2018ueo,Ogiya:2017hbr,Ishiyama:2019hmh,Colombi:2020xbv} --- a finding recently supported by novel analytical calculations~\cite{White:2022yoc,DelPopolo:2023tcp}. Intriguingly, studies indicate that these cusps, which follow a $\rho_{\text{cusp}}\propto r^{-1.5}$ radial density profile, form almost instantaneously \cite{Delos:2019mxl,Delos:2022yhn,Ondaro-Mallea:2023qat} and survive both halo growth \cite{Delos:2022yhn} and clustering \cite{Delos:2022bhp,Delos:2025pen}. 

Although dense cusps represent only a small fraction of the total dark matter content of the Universe, their high densities lead them to dominate the dark matter annihilation signature in models where dark matter can  annihilate~\cite{Blanco:2019eij, StenDelos:2019xdk, Delos:2022bhp, Delos:2023azx}. 
Consequently, prompt cusps can greatly boost the flux of Standard Model particles, including photons, charged cosmic rays, and neutrinos, produced by dark matter annihilation, while also significantly altering the spatial distribution of this flux. 
In regions where prompt cusps drive the annihilation rate, the annihilation morphology tracks the number of prompt cusps, effectively tracing the dark matter density, $\rho$. 
This contrasts with typical annihilation signals, 
which scale with the square of dark matter density, $\rho^2$, 
and instead more closely resembles the signal expected from decaying dark matter.

The similarity between annihilation in prompt cusps and dark matter decay motivates $\gamma$-ray observations of galaxy clusters. 
As the most massive structures in the Universe, galaxy clusters stand among the best targets for the indirect detection of dark matter decay~\cite{2010JCAP...05..025A, Huang:2011xr, 2012JCAP...07..017A, Tan:2019gmb, Thorpe-Morgan:2020czg, DiMauro:2023qat, Song:2023xdk}. Notably, galaxy cluster constraints on decaying dark matter can set lower-limits on the dark matter lifetime that exceed 10$^{27}$~s~\cite{Song:2023xdk}.

\begin{figure}[]
\centering
\includegraphics[width = 0.48\textwidth]{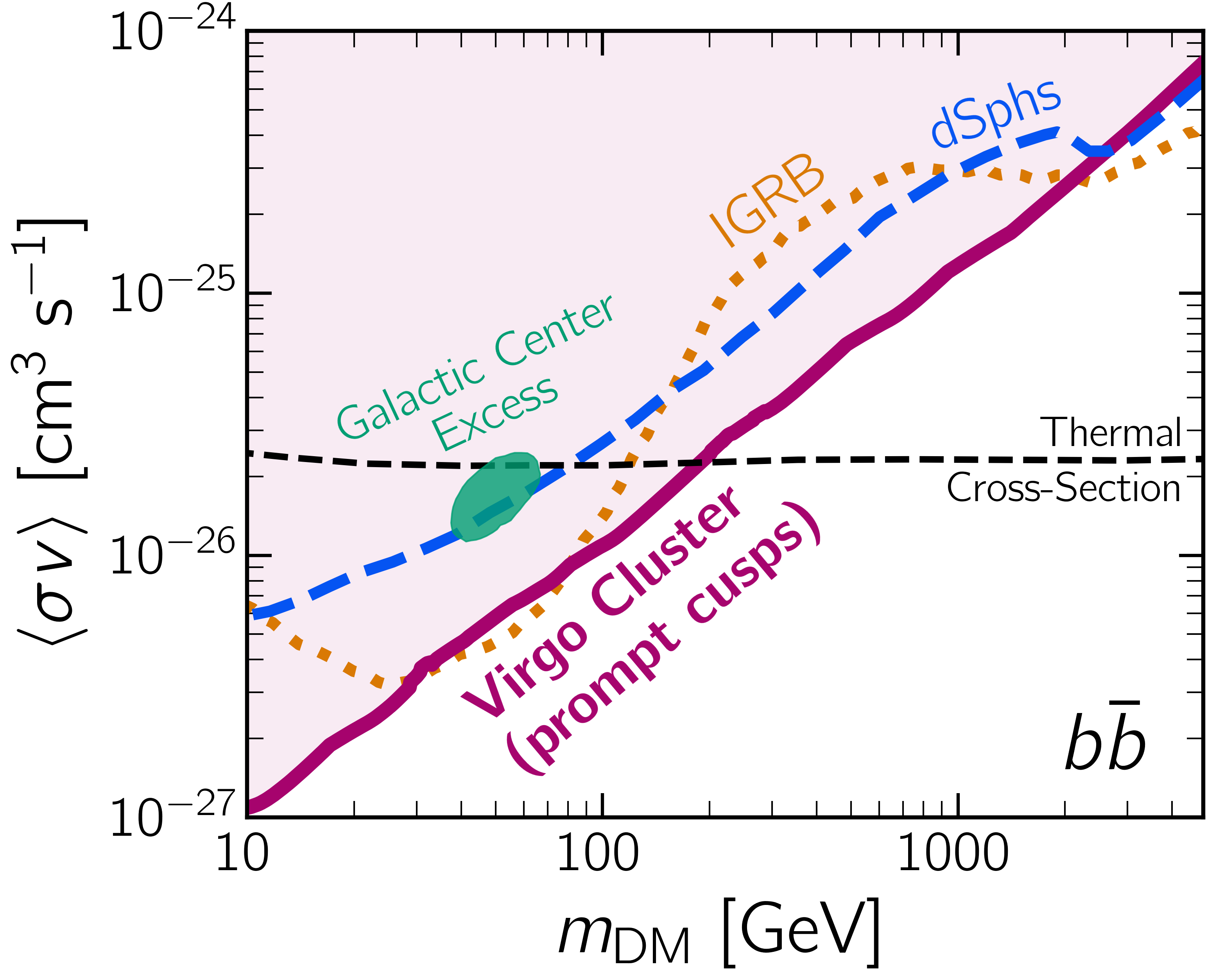}
\vspace{-0.5cm}
\caption{Upper limits on the dark matter annihilation cross-section to a $b\bar{b}$ final state in the Virgo cluster. Our limits (magenta solid) rule out the thermal annihilation cross-section (black dashed) for dark matter masses below 200~GeV~\cite{Steigman:2012nb}, exceeding constraints from the stacking of dwarf spheroidal galaxies~\cite{McDaniel:2023bju} (blue dashed) and from the IGRB \cite{Delos:2023ipo} (orange dashed). These results are in tension with dark matter explanations for the Galactic Center Excess (green region) \cite{Calore:2014xka}. The light-shaded region indicates the excluded parameter space.
}
\vspace{-0.3cm}
\label{fig:comparison}
\end{figure} 

Our analysis leverages the enhanced annihilation rate from prompt cusps to improve the sensitivity of $\gamma$-ray searches for dark matter. Most studies focus on $\gamma$-ray fluxes that scale with the square of dark matter density, favoring targets with high central densities, such as the Galactic Center \cite{Hooper:2010mq, Hooper:2011ti, Abazajian:2012pn, Zhou:2014lva, Daylan:2014rsa, Calore:2014nla, Calore:2014xka, Fermi-LAT:2017opo, Leane:2019xiy, Abazajian:2020tww, DiMauro:2021raz, Song:2024iup} or dwarf spheroidal galaxies (dSphs) ~\cite{Fermi-LAT:2010cni, Fermi-LAT:2013sme, Geringer-Sameth:2014qqa, Geringer-Sameth:2014yza, Fermi-LAT:2015att, Li:2015kag, Fermi-LAT:2016uux, Chiappo:2018mlt, Hoof:2018hyn, Li:2018kgy, Calore:2018sdx, Alvarez:2020cmw, Linden:2019soa, McDaniel:2023bju, Li:2021vqg, Crnogorcevic:2023ijs}. However, the annihilation rate from prompt cusps linearly depends on the total dark matter mass, elevating high mass targets (\emph{e.g.}, galaxy clusters or the isotropic $\gamma$-ray background (IGRB)~\cite{Delos:2023ipo}), to the most promising  candidates. 


In this \emph{paper}, we search for dark matter signals in the prompt cusps in galaxy clusters. First, we construct new models for the dark matter flux and morphology from seven nearby clusters, including Virgo, Centaurus, NGC 4636, M 49, Fornax (NGC 1399), Hydra, and Coma, accounting for both the internal density structure and spatial distribution of prompt cusps. We then analyze 15 years of $\gamma$-ray data from the \textit{Fermi} Large Area Telescope (\textit{Fermi}-LAT), finding no $\gamma$-ray excesses consistent with prompt cusp models. Figure~\ref{fig:comparison} shows that observations of Virgo alone set a constraint on the dark matter annihilation cross section that is significantly stronger than previous limits from dSphs and the IGRB.

The paper is organized as follows. In Section~\ref{sec:signal}, we characterize the annihilation signal from galaxy clusters, including the cosmological distribution of prompt cusps and the impact of tidal stripping. We then apply these models to specific clusters and compute the expected $\gamma$-ray flux. In Section~\ref{sec:data_analysis}, we present our data analysis methodology and the statistical techniques employed to search for the $\gamma$-ray annihilation signal. We present our results in Section~\ref{sec:results} and compare our findings with existing constraints. Finally, we conclude in Section~\ref{sec:concl}.

\section{Characterizing the annihilation signal}
\label{sec:signal}


The prompt cusps that dominate the annihilation signal form around redshift $z\sim 30$ \cite{Delos:2022bhp} and later accrete onto the galaxy cluster halos that we target. Our calculation of cluster halo annihilation signatures thus proceeds in two parts. First, we describe the average distribution of prompt cusps within an arbitrary volume of dark matter. Next, we specialize to clusters, accounting for their spatial structure and the disruptive effects that occur therein.

\subsection{Cosmological distribution of prompt cusps}
\label{sec:cusps}

Prompt cusps form from local maxima in the initial density field. Each of these density peaks has a collapse time $\ac$, which we estimate using the ellipsoidal collapse approximation in Ref.~\cite{Sheth:1999su}, and a characteristic comoving size $R\equiv|\delta/\nabla^2\delta|^{1/2}$, where $\delta$ is the height of the peak in the density contrast and $\nabla^2\delta$ is the Laplacian of the density contrast with respect to the comoving position. Numerical simulations \cite{Delos:2019mxl,Delos:2022yhn,Ondaro-Mallea:2023qat} show that a peak with these parameters collapses to form a power-law cusp with profile $\rho=A r^{-1.5}$, where:
\begin{align}\label{eq:Acusp}
    A&\simeq 24 \bar\rho(\ac)\,(\ac R)^{1.5}.
\end{align}
Here $\bar\rho(a)$ is the cosmic mean dark matter density at the scale factor $a$.
This profile initially extends out to the radius
\begin{align}\label{eq:rcusp}
    \Rc &\simeq 0.11 \ac R.
\end{align}
These prompt cusps grow halos around them over time, and the halos also contribute to dark matter annihilation.
However, prior halo and subhalo modeling suggests that their contribution to the annihilation rate is only at about the 10 percent level \cite{Delos:2022bhp}, so we neglect them. We will revisit this comparison in Section~\ref{sec:clusters}.

Since the annihilation rate arising from a $\rho\propto r^{-1.5}$ density cusp diverges at small radii, it is also important to understand the deep interior structure of a prompt cusp.
Analytic arguments \cite{Delos:2022yhn} (approximately confirmed in simulations \cite{Maccio:2012qf}) require that the initial thermal motion of the dark matter give rise to a finite-density core of radius
\begin{align}\label{eq:rcore}
    \rc \simeq 0.34 G^{-2/3} (\mDM/\Td)^{-2/3} \bar\rho(\ad)^{-4/9}A^{-2/9},
\end{align}
where $\mDM$ is the dark matter particle mass, $\Td$ is the temperature at which the dark matter kinetically decouples from the Standard Model plasma, and $\bar\rho(\ad)$ is the dark matter density at the time of kinetic decoupling \cite{Delos:2022bhp}. Note that typically $\rc\sim 10^{-3}\Rc$.
Assuming that the $\rho=Ar^{-1.5}$ density profile transitions abruptly into a constant-density core at the radius $\rc$, the annihilation rate from radii below $\Rc$ is proportional to the quantity
\begin{align}\label{eq:Jcusp}
    j \equiv \int_\mathrm{cusp}\rho^2 \diff V = 4\pi A^2[1/3+\ln(\Rc/\rc)].
\end{align}
More realistic models of the central core lead to a slightly higher annihilation rate \cite{Stucker:2023rjr}.

The cosmological distribution of prompt cusps follows directly from the distribution of peaks in the initial density field, which is a well understood mathematical problem \cite{Bardeen:1985tr}.
To describe the initial density field, we fix a dark matter power spectrum using the procedure described by Refs.~\cite{Delos:2022bhp,Delos:2023azx,Delos:2023ipo}. We begin by evaluating the dark matter power spectrum at $z=31$ using \textsc{class} \cite{Blas:2011rf} with cosmological parameters from the Planck mission \cite{Planck:2018vyg}, and we extrapolate it beyond the code's numerical resolution limit using the analytic solution of Ref.~\cite{Hu:1995en} that is valid in the small-scale limit.
The result is the spectrum of initial density variations for an idealized cold dark matter model, but in practice, the thermal motion of the dark matter suppresses inhomogeneities on smaller scales than the distances that particles randomly stream.
The free-streaming scale depends on the dark matter model, namely its mass $\mDM$ and decoupling temperature $\Td$ (which is set by its elastic interactions with the Standard Model).
We consider a range of models, and for each one, we evaluate the free-streaming wavenumber $\kfs$ according to Ref.~\cite{Bertschinger:2006nq} and scale the power spectrum by $\e^{-k^2/\kfs^2}$.
Prompt cusps arise from density perturbations with length scales of order $\kfs^{-1}$.

With the power spectrum set, we follow the approach of Ref.~\cite{Delos:2022bhp} to evaluate the statistics of initial peaks and to translate this into the cosmological abundance and distribution of prompt cusps.
The procedure is to evaluate the comoving number density $n_\mathrm{peaks}$ of initial peaks and to randomly sample the properties of $10^7$ of them.
Any mass $M$ of dark matter contains an average of $M n_\mathrm{peaks}/\bar\rho_0$ initial peaks, where $\bar\rho_0\simeq 33$~M$_\odot$kpc$^{-3}$ is the comoving dark matter density.
If we associate each peak with a cusp that contributes some $j$ to the annihilation rate, in accordance with Eq.~(\ref{eq:Jcusp}), then these cusps would contribute in total
\begin{align}
    \int_\mathrm{cusps}\rho^2\diff V
    =M n_\mathrm{peaks}\langle j\rangle/\bar\rho_0,
\end{align}
where angle brackets indicate the average.
The average contribution of the prompt cusps per mass of dark matter is:
\begin{equation}\label{eq:rhoeff0}
    \rhoeffo
    \equiv \frac{\int_\mathrm{cusps}\rho^2\diff V}{M}
    =n_\mathrm{peaks}\langle j\rangle/\bar\rho_0.
\end{equation}
For a dark matter particle of mass $\mDM$ and velocity-weighted annihilation cross-section $\sigmav$ (which we assume to be velocity-independent at lowest order), the cusps would contribute an average annihilation rate
\begin{align}\label{eq:GoverM0}
    \left\langle\frac{\Gamma}{\MDM}\right\rangle_{\!0}
    &=
    \frac{\sigmav}{2m_\mathrm{DM}^2}\rhoeffo
\end{align}
per mass of dark matter.

So far, we have assumed that each initial peak can be associated with a prompt cusp with density profile given by Eqs~(\ref{eq:Acusp}-\ref{eq:rcore}), which is why we use the subscript `0' in Eqs. (\ref{eq:rhoeff0}) and~(\ref{eq:GoverM0}).
This assumption does not fully hold.
Some initial peaks do not collapse to form prompt cusps because they accrete into previously formed structures first. Additionally, some of the prompt cusps that do form are destroyed in halo mergers. The approximate picture is that when a small halo accretes onto a much larger halo, the small halo persists as a subhalo, and the central cusps of both halos survive. On the other hand, when halos of comparable mass merge, dynamical friction causes their central cusps to also merge, and this process decreases the number of prompt cusps (and the inner structure of the central cusp of the larger halo is only slightly altered \cite{Delos:2022yhn}).
Based on the simulations of Ref.~\cite{Delos:2022yhn}, Ref.~\cite{Delos:2022bhp} estimated that due to these effects, only about half of the initial density peaks produce prompt cusps that persist today.
These surviving cusps contribute a fraction $\gfac$ of the annihilation signal predicted by Eq.~(\ref{eq:rhoeff0}), where $\gfac\simeq 0.5\pm0.1$.
We will adopt $\gfac=0.5$.

\subsection{Annihilation in galaxy clusters}
\label{sec:clusters}

Now, we consider a cluster halo with density profile $\rho(r)$.
Let $\diff N_\gamma/\diff E$ be the spectrum of $\gamma$-ray photons produced by each dark matter annihilation event.
The differential photon flux (per solid angle) received by an observer is then
\begin{align}\label{eq:flux0}
	\frac{\diff^2\Phi}{\diff\Omega \diff E}&=
	\frac{1}{4\pi}
	\frac{\diff N_\gamma}{\diff E}
	\int_{\text{l.o.s.}}\diff\ell\,
	\rho(r)
    \left\langle\frac{\Gamma}{\MDM}\right\rangle,
\end{align}
given in terms of the annihilation rate per mass of dark matter, $\langle\Gamma/\MDM\rangle$.
Here, $\ell$ is the position along the observer's line of sight, the cluster-centric radius $r$ accordingly depends on $\ell$, and the line-of-sight integration is performed out to the virial radius of each cluster.
We expand the annihilation rate as
\begin{align}\label{eq:annihilationrate}
    \left\langle\frac{\Gamma}{\MDM}\right\rangle 
    &=
    \frac{\sigmav}{2m_\mathrm{DM}^2}\left[\rho(r)+\gfac\lfac(r)\rhoeffo\right].
\end{align}
The first term represents the smooth halo~\cite{Salucci:2018hqu}, which contains almost all of the dark matter, as only about 1\% resides in prompt cusps \cite{Delos:2022bhp}.
The second term accounts for the cusps and is constructed by scaling Eq.~(\ref{eq:GoverM0}) by the cusp survival factor $\gfac$ and by a position-dependent factor $\lfac(r)$.

\begin{table*}
\centering
    \begin{ruledtabular}
    \begin{tabular}{ccccccccc}
    Name & RA, Dec (deg.) & distance (Mpc) & $\rho_s$ (M$_\odot$Mpc$^{-3}$) & $r_\mathrm{s}$ (Mpc) & $R_{200}$ (Mpc) & $\theta_{200}$ & $M_{200}$ (M$_\odot$) & $J$ (GeV$^2$\,cm$^{-5}$) \\
    \hline
    Virgo & 187.70, 12.34 & $15.46$ & $1.22 \times 10^{15}$ & $0.335$ & $1.700$ & $6.3^{\circ}$ & $5.6 \times 10^{14}$ & $8.88 \times 10^{20}$ \\
    NGC 4636 & 190.71, 2.69 & $17.18$ & $1.64 \times 10^{15}$ & $0.135$ & $0.777$ & $2.6^{\circ}$ & $5.3 \times 10^{13}$ & $6.85 \times 10^{19}$ \\
    Hydra & 139.53, $-12.09$ & $47.51$ & $1.3 \times 10^{15}$ & $0.264$ & $1.376$ & $1.7^{\circ}$ & $3.0 \times 10^{14}$ & $4.99 \times 10^{19}$ \\
    Coma & 194.95, 27.98 & $100.24$ & $1.17 \times 10^{15}$ & $0.454$ & $2.260$ & $1.3^{\circ}$ & $1.3 \times 10^{15}$ & $4.95 \times 10^{19}$ \\
    M49 & 187.44, 8.00 & $18.91$ & $1.68 \times 10^{15}$ & $0.127$ & $0.741$ & $2.2^{\circ}$ & $4.6 \times 10^{13}$ & $4.88 \times 10^{19}$ \\
    Centaurus & 192.20, $-41.31$ & $43.16$ & $1.33 \times 10^{15}$ & $0.239$ & $1.258$ & $1.7^{\circ}$ & $2.3 \times 10^{14}$ & $4.61 \times 10^{19}$ \\
    Fornax & 54.62, $-35.45$ & $21.50$ & $1.65 \times 10^{15}$ & $0.132$ & $0.763$ & $2.0^{\circ}$ & $5.1 \times 10^{13}$ & $4.12 \times 10^{19}$ \\
    \end{tabular}
    \end{ruledtabular}
\caption{
\label{tab:clusters}Clusters considered in our analysis and the properties that we adopt. For each cluster, we adopt a Navarro-Frenk-White (NFW) density profile, $\rho(r)=\rho_\mathrm{s}(r/r_\mathrm{s})^{-1}(1+r/r_\mathrm{s})^{-2}$ \cite{Navarro:1995iw,Navarro:1996gj}, extending out to the virial radius $R_{200}$, which encloses mass $M_{200}$ and corresponds to an angle $\theta_{200}$ on the sky. Distances and structural parameters are drawn from Ref.~\cite{DiMauro:2023qat}. We also show the integrated annihilation $J$ factor calculated in accordance with Section~\ref{sec:signal}. $J$ depends mildly on the dark matter mass and decoupling temperature; here we fix $\mDM=100$~GeV and $\Td=30$~MeV.}
\end{table*}

\begin{figure}[tbp]
\centering
\includegraphics[width=\columnwidth]{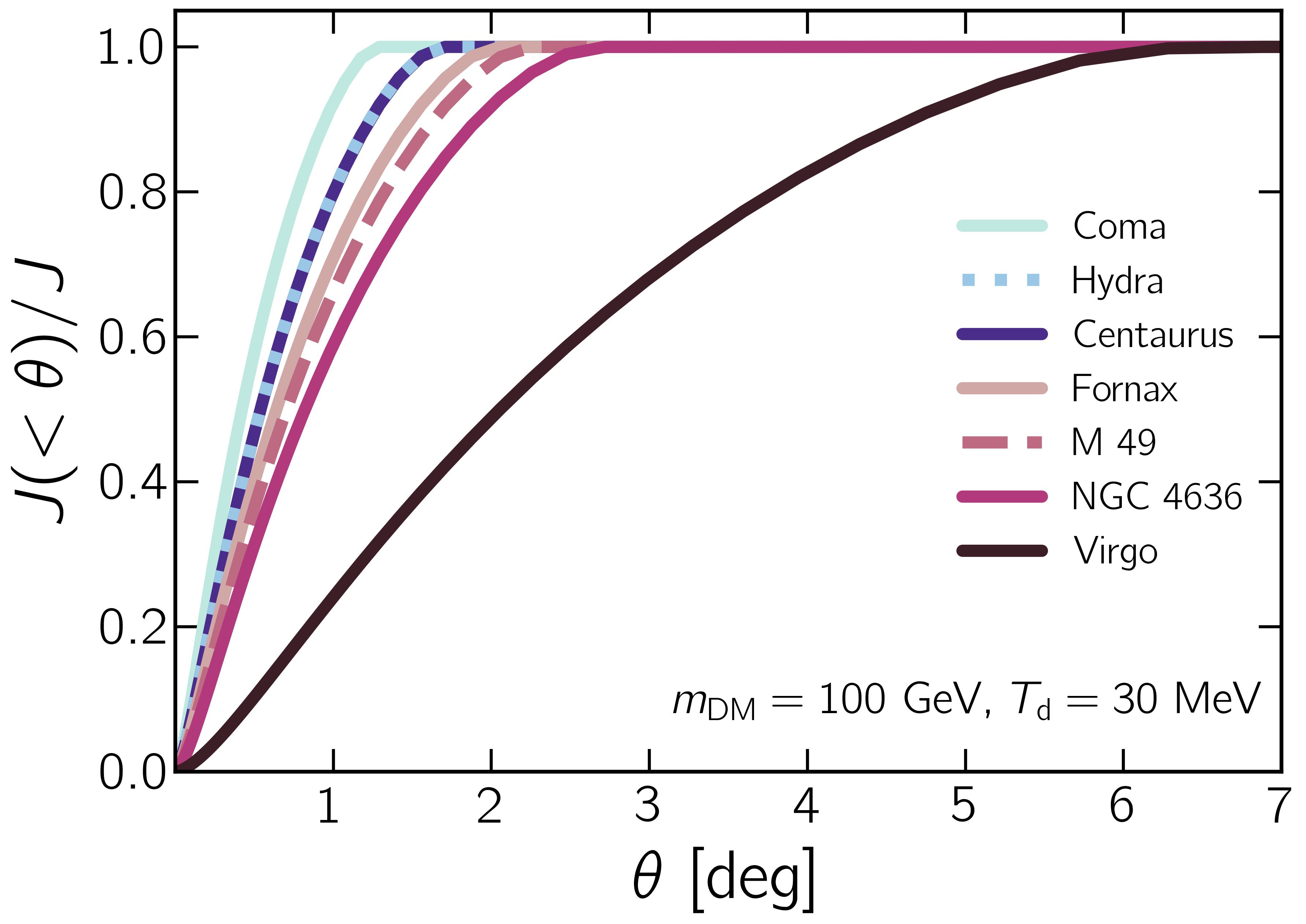}
\caption{
Spatial extent of each cluster's annihilation signal. For each angle $\theta$ from the cluster center, we show the fractional contribution to the $J$ factor that comes from angles less than $\theta$.
}
\label{fig:cluster-J-angle}
\end{figure}

Inside the halo of a galaxy cluster, prompt cusps are gradually stripped by tidal forces.
To account for this effect, we employ the model of Ref.~\cite{Stucker:2022fbn}, which uses the tidal field at a subhalo's orbital pericenter to predict the density profile of its asymptotic remnant after an arbitrarily long period of tidal stripping.
For cusps at a radius $r$ within a cluster, we make a simple approximation that the pericenter radii $r_\mathrm{p}$ are uniformly distributed between $r_\mathrm{p}=0$ and $r_\mathrm{p}=r$. Appendix~\ref{sec:peri} shows that this is a reasonable approximation.
For each cusp sampled in Section~\ref{sec:cusps}, which had the initial annihilation rate coefficient $j$ given by Eq.~(\ref{eq:Jcusp}), we can evaluate a new coefficient $j^\prime=\int\tilde \rho^2\diff V$ associated with the cusp's tidally evolved density profile $\rho^\prime$. We then define
\begin{align}
    \lfac(r) = \langle j^\prime\rangle/\langle j\rangle,
\end{align}
where the brackets average over the pericenter radii $r_\mathrm{p}<r$.

We consider the clusters in Table~\ref{tab:clusters} and adopt the density profiles $\rho(r)$ and distances from Ref.~\cite{DiMauro:2023qat}.
From Eqs. (\ref{eq:flux0}) and~(\ref{eq:annihilationrate}), the differential photon flux from each cluster is:
\begin{align}\label{eq:flux}
	\frac{\diff^2\Phi}{\diff\Omega \diff E}&=
	\frac{\sigmav}{8\pi m_\mathrm{DM}^2}
	\frac{\diff N_\gamma}{\diff E}
    \frac{\diff J}{\diff\Omega},
\end{align}
where the spatial distribution of the dark matter is in the factor
\begin{align}\label{eq:dJdO}
    \frac{\diff J}{\diff\Omega}\equiv
	\int_{\text{l.o.s.}}\diff\ell\,
	\rho(r)
    [\rho(r)+\gfac\lfac(r)\rhoeffo].
\end{align}
Integrated over solid angles, the total annihilation luminosity of a cluster is proportional to
\begin{align}
    J = \int\diff\Omega\frac{\diff J}{\diff\Omega}.
\end{align}
Table~\ref{tab:clusters} gives the annihilation $J$ factor of each cluster (integrated out to the virial radius) for an example model with $\mDM=100$~GeV and $\Td=30$~MeV. Figure~\ref{fig:cluster-J-angle} shows the annihilation morphology over the sky. Due to its proximity, the Virgo cluster has the highest $J$ by far, but it also produces the most spatially extended signal.

\begin{figure}[tbp]
\centering
\includegraphics[width=\columnwidth]{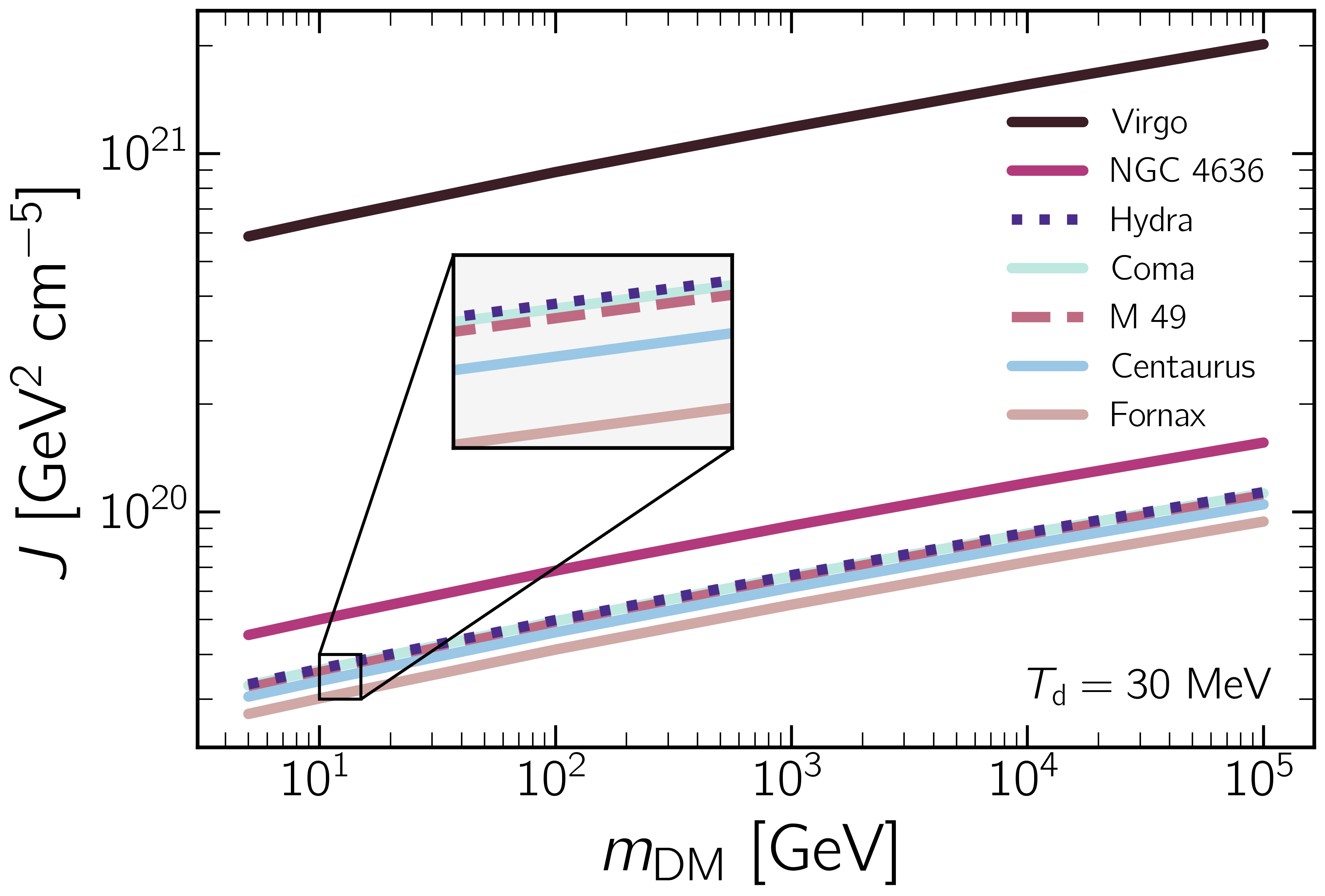}
\caption{
Cluster $J$ factors, shown as a function of the dark matter mass.
Higher particle masses correspond to colder dark matter, so the prompt cusps are smaller and form at earlier times. The earlier formation makes these cusps more internally dense, resulting in a higher annihilation rate.
}
\label{fig:cluster-J}
\end{figure}

Generally, higher dark matter masses and earlier decoupling (higher $\Td$) both lead to larger values of $J$, because they cause the prompt cusps to be slightly more internally dense. However, $J$ depends only logarithmically on $\mDM$ and $\Td$ \cite{Delos:2022bhp}.
We follow previous studies \cite{Fermi-LAT:2015qzw,Delos:2023ipo} in fixing $\Td=30$~MeV for simplicity.
This is a mildly conservative choice; lower $\Td$ tend to be in tension with terrestrial experiments \cite{Cornell:2013rza}, since they require a strong kinetic coupling between the dark matter and the Standard Model.
We however analyze a range of dark matter masses $\mDM$, and Fig.~\ref{fig:cluster-J} shows how each cluster's $J$ factor varies with $\mDM$ (with $\Td=30$~MeV fixed). The spatial distribution of the annihilation signal also depends in principle on $\mDM$, since denser cusps are less susceptible to tidal stripping. However, this variation is small enough that Fig.~\ref{fig:cluster-J-angle} does not discernibly change over the full range of $5~\text{GeV}<\mDM<10^5~\text{GeV}$.

\begin{figure}[tbp]
\centering
\includegraphics[width=\columnwidth]{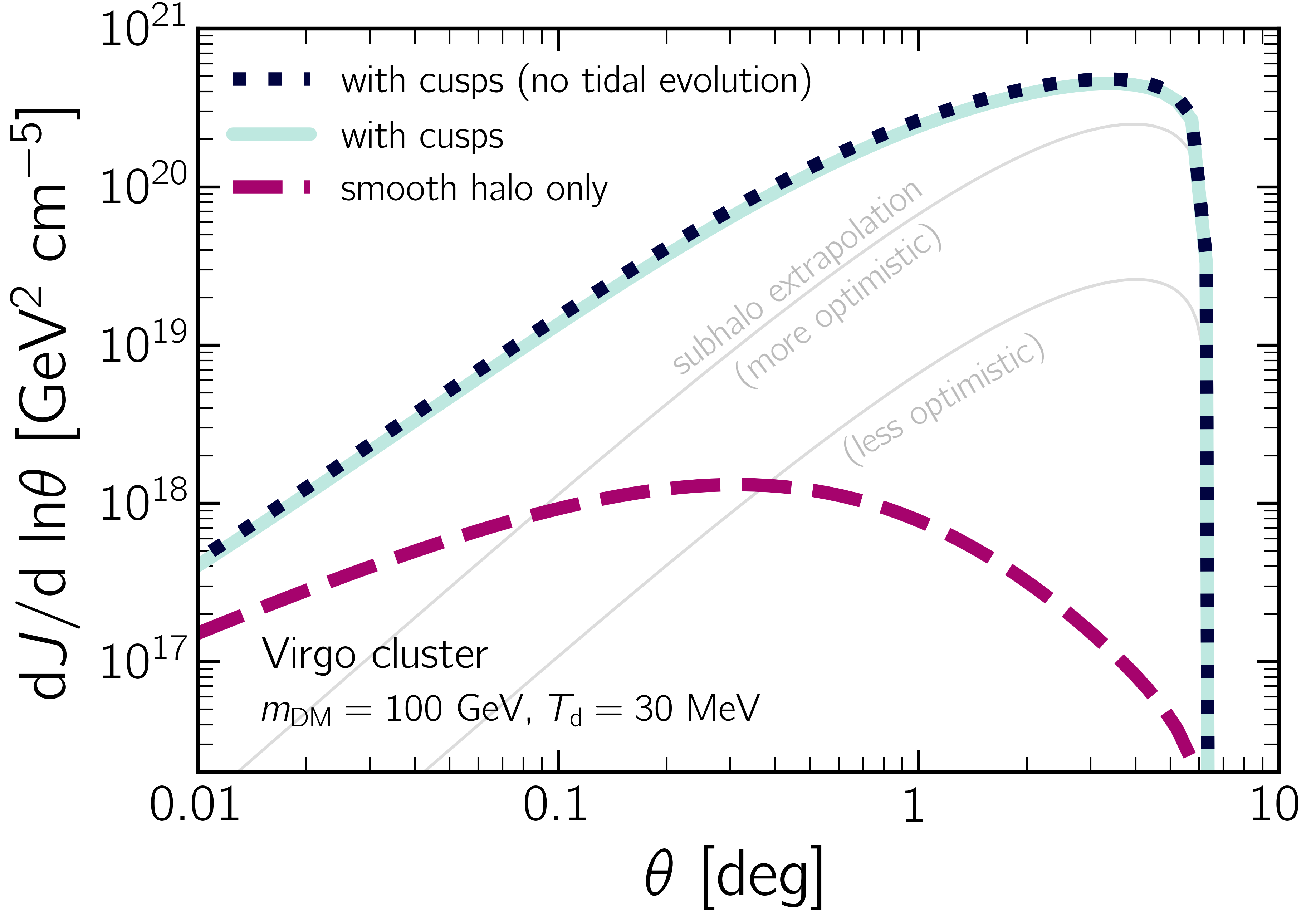}
\caption{
Comparing different contributions to the annihilation signal of the Virgo cluster.
As a function of the angle $\theta$ from the cluster center, we show the contribution $\diff J/\diff\ln\theta = 2\pi\theta^2 \diff J/\diff\Omega$ of each logarithmic angular interval to the integrated $J$ factor.
The light teal solid curve shows the net predicted $J$ factor, which we use for our analysis.
The navy dotted curve shows the $J$ factor without accounting for tidal stripping of prompt cusps (5--10\% higher). The magenta dashed curve shows the $J$ factor from the smooth halo alone, neglecting contributions from cusps altogether.
For comparison, the two gray lines show the contribution from subhalos (without prompt cusps) estimated according to two commonly used extrapolations (see the text).
}
\label{fig:J-breakdown}
\end{figure}

Figure~\ref{fig:J-breakdown} illustrates the different contributions to the $J$ factor in Eq.~(\ref{eq:dJdO}). The total signal from the smooth halo is subdominant compared to the signal from the prompt cusps and only contributes nonnegligibly at angles smaller than about 0.01$^\circ$ from the cluster center. Such angles are far smaller than the angular resolution of the Fermi telescope.
Tidal forces reduce the annihilation signal from the cusps by only about 5--10\%.

Previous efforts to constrain dark matter annihilation using galaxy clusters and large-scale structure (e.g.~\cite{Fermi-LAT:2015xij,Chan:2017spr,Bartlett:2022ztj,DiMauro:2023qat,Manna:2024dna}) have primarily relied on the contribution from unresolved subhalos (without accounting for their central prompt cusps). Therefore, although in this work we neglect the contribution of subhalos beyond the prompt cusps, it is instructive to include a comparison in Fig.~\ref{fig:J-breakdown}. Here we show two estimates of the angular distribution of the $J$ factor that result from standard subhalo models. We consider top-level subhalos, subhalos of those subhalos, and so on up to 6 levels of nesting; we verified that further iteration does not change the results. Within each host, subhalos are spatially distributed in accordance with the number density distribution obtained by Ref.~\cite{Springel:2008cc}. The density profile of each subhalo is NFW, set by the subhalo concentration-mass relation from Ref.~\cite{Moline:2016pbm}. Finally, the number of subhalos of each mass $m$ is taken to be $\diff N/\diff m = (A_\mathrm{sub}/M)(m/M)^{-\alpha_\mathrm{sub}}$ in a host of mass $M$, ranging from a maximum of $m=0.01M$ to a minimum of $m=10^{-6}$~M$_\odot$ (approximately the mass of a prompt cusp for dark matter with $\mDM=100$~GeV and $\Td=30$~MeV \cite{Delos:2022bhp}).
We set $(A_\mathrm{sub},\alpha_\mathrm{sub})=(0.012,2)$ for a more optimistic estimate and $(A_\mathrm{sub},\alpha_\mathrm{sub})=(0.03,1.9)$ for a less optimistic estimate \cite{Sanchez-Conde:2013yxa}.
These estimates are similar to the ``MAX'' and ``MED'' subhalo models from Ref.~\cite{DiMauro:2023qat} and result in subhalos contributing 8.5 and 83 times the integrated annihilation luminosity of the smooth halo of the Virgo cluster, respectively.
In comparison, the cusps contribute integrated luminosity about 200 times that of the smooth halo.

The estimated subhalo contribution shown in Fig.~\ref{fig:J-breakdown} is significantly more spatially extended than the contribution from the cusps. This outcome is due to tidal stripping of the subhalos (e.g.~\cite{Han:2015pua}). While prompt cusps are also affected by tidal stripping, their more centrally concentrated structures make them more resistant to tidal effects \cite{Penarrubia:2010jk,Du:2024sbt,Aguirre-Santaella:2025oyz} while also suppressing the impact of any mass loss on the annihilation luminosity \cite{Delos:2022bhp}.
At large $\theta$, the more optimistic subhalo estimate lies within a factor of 2 of the cusp contribution, while the less optimistic estimate is lower by at least an order of magnitude at all $\theta$.
Note, however, that the subhalo estimates are based on simulations of roughly $10^5$~M$_\odot$ to $10^{10}$~M$_\odot$ subhalos of $10^{12}$~M$_\odot$ field halos \cite{Diemand:2008in,Springel:2008cc}. Calculating the annihilation signal from subhalos involves extrapolating these results by more than 10 orders of magnitude in both subhalo mass and host halo mass (for nested subhalos).
Consequently, these estimates are highly uncertain (see also Refs.~\cite{Okoli:2017pqr,Drakos:2023mih}). Indeed, a more recent semianalytic subhalo modeling approach found that subhalos contribute less annihilation luminosity than both of the extrapolations we have presented \cite{Hiroshima:2018kfv,Ando:2019xlm}.
It is because of these uncertainties that we do not include the contribution from the subhalos (beyond their central prompt cusps) in our analysis.

\subsection{Robustness of the predicted signals}

We emphasize that the annihilation signal from the prompt cusps is much clearer than that coming from unresolved subhalos.
The initial distribution of cusps emerges directly from the statistics of the cosmological initial conditions \cite{Delos:2019mxl,Delos:2022bhp}, and tidal stripping during their evolution as substructures can only marginally reduce the annihilation luminosity from the cusps \cite{Delos:2022bhp}.
The one extrapolation that we make involves the number of cusps that survive the hierarchical assembly of cosmic structure (as discussed at the end of Section~\ref{sec:cusps}).  Reference~\cite{Delos:2022bhp} found that the annihilation rate contributed by surviving cusps rapidly converges, becoming essentially constant by a time that corresponds to $z\simeq 7$ in our scenarios.\footnote{Reference~\cite{Delos:2022bhp} found that the annihilation rate is constant by the time that $\sigma_0\simeq 3$, where $\sigma_0$ is the rms fractional density contrast in linear theory. For dark matter with $\mDM=100$~GeV and $\Td=30$~MeV, that happens at redshift $z\simeq 7$.} We assume that this contribution remains constant thereafter. This assumption is well justified, however, because loss of prompt cusps through dynamical friction inspirals should become rare when typical halos are much more massive than the cusps.
Cusps can also be disrupted by close encounters with stars, but this effect is only significant near the dense centers of galaxies \cite{Stucker:2023rjr}.

There are nevertheless two main modeling uncertainties in the signal from the prompt cusps.
First, the numerical factor in Eq.~(\ref{eq:Acusp}), by which cusp coefficients $A$ are evaluated, is uncertain by around 20 percent. We chose the prefactor 24 suggested by Ref.~\cite{Delos:2022yhn}, but Refs.~\cite{Delos:2019mxl,Ondaro-Mallea:2023qat} obtained somewhat different prefactors of 26.5 and 19.1, respectively. This 20 percent uncertainty in $A$ translates to about a factor of 1.5 uncertainty in the annihilation rate.
Second, Ref.~\cite{Delos:2022bhp} found that the fraction $\gfac$ of the total cusp annihilation luminosity that survives hierarchical assembly is uncertain by about 20 percent.
Overall, there is close to a factor of 2 uncertainty in the annihilation signal from the cusps.
We follow Ref.~\cite{Delos:2023ipo} in not incorporating this uncertainty into our results. Since it propagates cleanly through our analyses, the annihilation limits that we derive can be straightforwardly modified to account for any future update to $\gfac$ or the numerical prefactor in Eq.~(\ref{eq:Acusp}).
Although this level of uncertainty is significant, we note that conventional annihilation searches using dwarf spheroidal galaxies can face similar or worse levels of uncertainty arising from incomplete knowledge about how the mass is distributed in these systems (e.g.~\cite{Sanders:2016eie,Hayashi:2016kcy,Chang:2020rem}). The signal from prompt cusps does not face that uncertainty because it depends primarily on the number of cusps, and hence only on the total mass of the dark matter system.


Also, the signal from prompt cusps is sensitive to the cosmological initial conditions on far smaller scales than are constrained by observations. As we discussed in Section~\ref{sec:clusters}, the annihilation rate scales logarithmically with the decoupling temperature $\Td$, since $\Td$ influences the dark matter free-streaming scale.\footnote{Reference~\cite{Delos:2022bhp} found that the annihilation rate scales as
\begin{equation}
    \rhoeffo\simeq 0.08\left[\ln\!\left(\frac{\mDM}{\mathrm{GeV}}\frac{\Td}{\mathrm{GeV}}\right)+36\right]^5\rhoc.
\end{equation}
}
Our choice to fix $\Td=30$~MeV is mildly conservative (see Section~\ref{sec:clusters}), and dark matter with higher $\Td$ could yield an annihilation signal that is more luminous by a factor of a few.
Moreover, the prompt cusp signal is sensitive to the small-scale matter power spectrum more broadly, and the annihilation signal could be greatly altered if either the primordial power spectrum \cite{Delos:2017thv,Delos:2018ueo} or the early thermal history of the Universe \cite{StenDelos:2019xdk,Delos:2023vfv,Ganjoo:2024mie,Bae:2025uqa,Fairbairn:2025cae} deviates from the simplest assumptions.
Note that these considerations are not specific to prompt cusps, and they are also relevant to the signal from unresolved subhalos used by many previous annihilation searches (e.g.~\cite{Fermi-LAT:2015xij,Fermi-LAT:2015qzw,Liu:2016ngs,Chan:2017spr,Bartlett:2022ztj,DiMauro:2023qat,Manna:2024dna,Cholis:2024hmd}).




\section{Data Analysis}
\label{sec:data_analysis}

The Large Area Telescope (LAT), onboard the \textit{Fermi} $\gamma$-ray observatory, detects $\gamma$ rays with energies spanning 20~MeV-- 500~GeV. Its field of view (FoV) covers $\sim$20\% of the sky at all times, allowing full-sky coverage at a cadence of about 3 hours. Its source localization accuracy reaches a few arcminutes. The instrument paper  provides a comprehensive overview of the LAT instrument and its performance~\cite{Atwood_2009}.

We analyze 15 years of {\tt Pass 8} reprocessed data covering the period from August 4, 2008, to November 1, 2023, obtained from the \textit{Fermi} Science Support Center (FSSC) website\footnote{\url{https://fermi.gsfc.nasa.gov/ssc/data/}, accessed on January 19, 2025.}. All data are analyzed using the {\tt fermipy} software package (v.~1.2.0) \cite{fermipyopensourcepythonpackage}, which relies on {\tt FermiTools} (v.~2.2.0)\footnote{\url{https://fermi.gsfc.nasa.gov/ssc/data/analysis/software/}, accessed on January 19, 2025.} for likelihood-based modeling and fitting. We select photons from the {\tt P8R3\_SOURCE\_V3} event class with energies between 500~MeV and 500~GeV, spaced in 8 logarithmic bins in each decade in energy. We apply the standard models for the Galactic diffuse emission ({\tt gll\_iem\_v07}) and the isotropic diffuse background ({\tt iso\_P8R3\_SOURCE\_V3}), as provided by the \textit{Fermi}-LAT Collaboration. For each source, we define a 20$^\circ$$\times$20$^\circ$ region of interest (RoI), binned at 0.2$^\circ$ and centered on the cluster's dynamical center. We model both the point and extended sources within the RoI using the 4FGL-DR4 catalog ({\tt gll\_psc\_v31}), extending to 25$^\circ$$\times$25$^\circ$ to account for photon leakage from the LAT point spread function. We introduce a zenith angle cut of 90$^\circ$ to account for possible contamination from the Earth's limb and apply the standard good time interval (GTI) selection cuts. 


\begin{figure*}[htbp]
\centering

\includegraphics[width=0.98\textwidth]{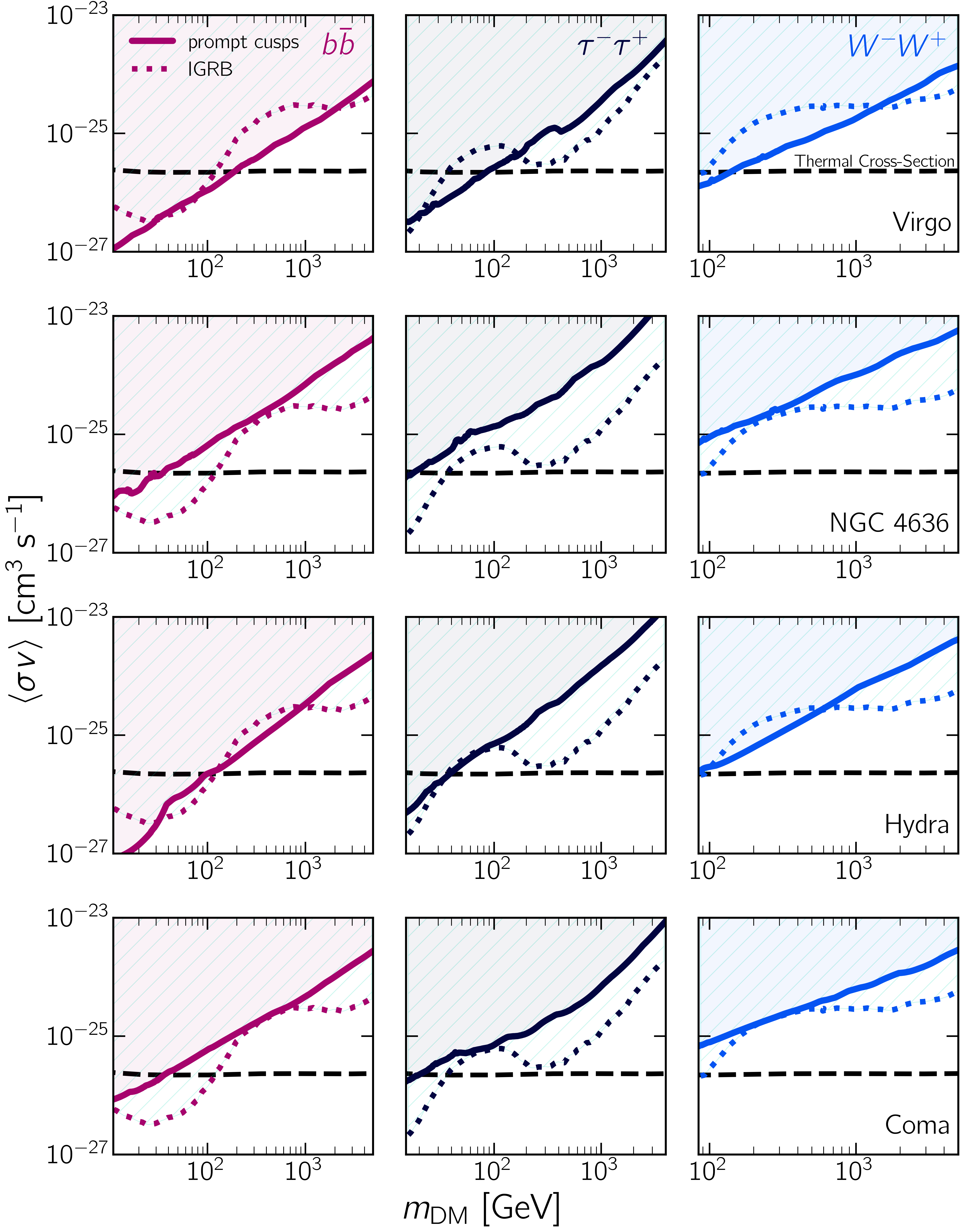} 
\caption{Upper limits on the annihilation cross-section from our seven selected galaxy clusters as a function of dark matter mass,
based on the extended source models
discussed in Section~\ref{sec:signal}. We consider three annihilation channels: $b\bar{b}$ (magenta),  $\tau^+\tau^-$ (navy), and $W^+W^-$ (blue). We compare our results (solid lines) with previous upper limits from the studies of the IGRB (dotted lines),  from Ref.~\cite{Delos:2023ipo}. The canonical thermal annihilation cross-section is shown as black dashed line~\cite{Steigman:2012nb}. The light-shaded regions represent the parameter space excluded by galaxy cluster observations, while the hatched regions indicate exclusions based on the IGRB analysis. We find that Virgo constraints are typically the strongest, ruling out dark matter annihilating at the thermal cross-section for masses up to 200~GeV, 88~GeV, and 147~GeV for annihilations to $b\bar{b}$, $\tau^+\tau^-$, and $W^+W^-$, respectively. 
}
\label{fig:ul_clusters}
\end{figure*}

\renewcommand\thefigure{5 continued}

\begin{figure*}
\includegraphics[width=0.98\textwidth]{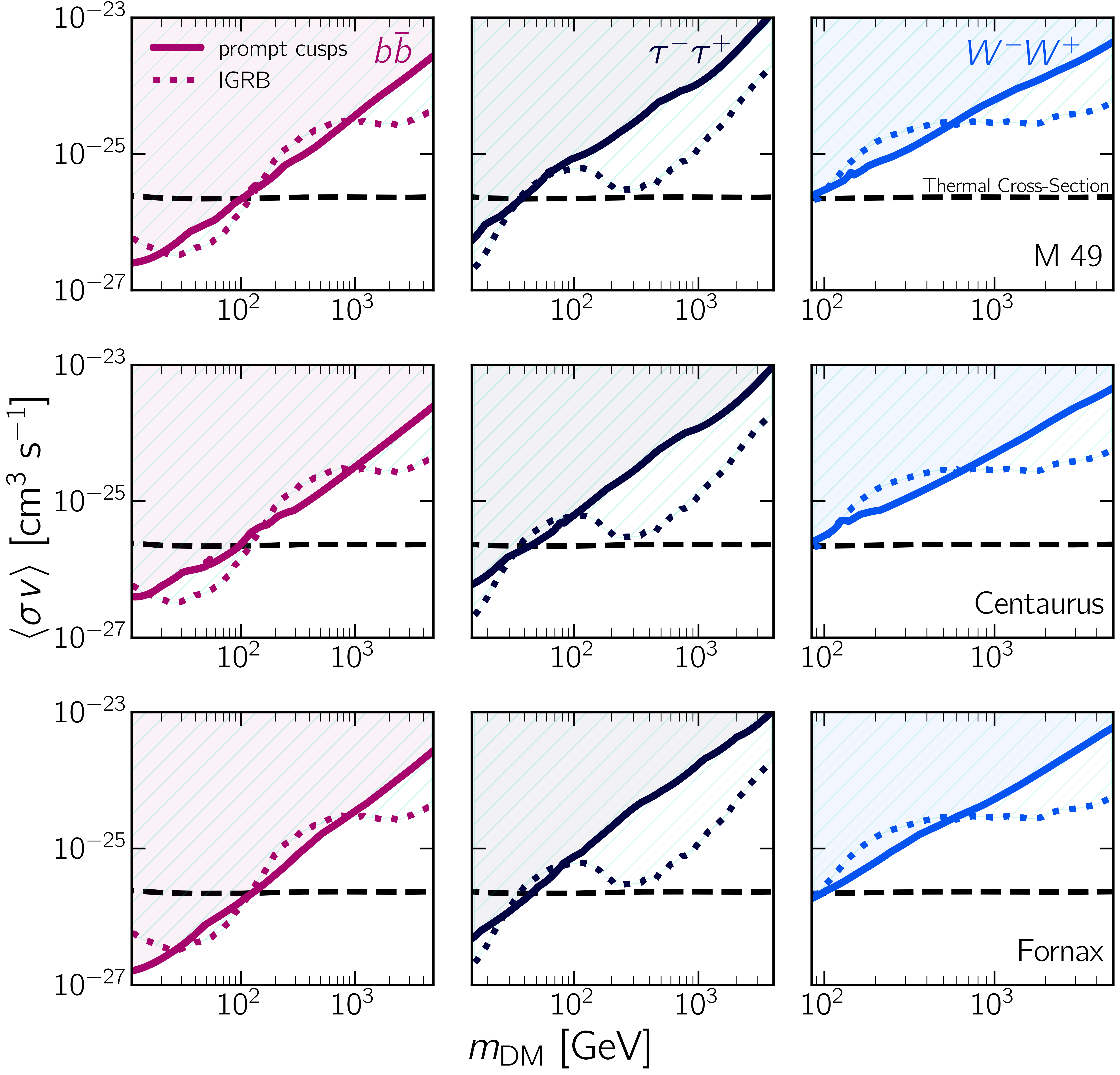}
\caption{}
\label{fig:ul_clusters2}
\end{figure*}
\renewcommand\thefigure{\arabic{figure}}
\setcounter{figure}{5}

After constructing the model, we perform a binned likelihood analysis for each source. The likelihood function compares models with and without an associated $\gamma$-ray source, quantifying the goodness of fit using the Test Statistic (TS), defined as $TS = -2~{\rm log} \left(\mathcal{L}_0/\mathcal{L}_1\right)$, where $\mathcal{L}_0$ represents the likelihood of the null hypothesis (no signal), and $\mathcal{L}_1$ is the likelihood of the alternative hypothesis (signal present). We perform an extended-source analysis using the dark-matter signal templates described in Sec.~\ref{sec:signal} for the seven clusters from Table \ref{tab:clusters}. We model the dark matter profile from each cluster out to 6$^\circ$ from the cluster center, in order to model the entire dark matter contribution (see Figure~\ref{fig:cluster-J-angle}). For each cluster, we use the \texttt{gta.sed()} module in \texttt{fermipy} to estimate the source flux by fitting the spectrum in each energy bin assuming a power-law shape with an index of 2. This method also produces a spectral energy distribution (SED) curve, and energy flux $E^2 \diff N /\diff E$, along with a corresponding delta log-likelihood profile, $\Delta \mathcal{L}\left(\diff \Phi_{\gamma}/\diff E, E\right)$. 

Once the flux (or upper limit) is determined, we translate the SEDs into dark matter constraints. The $\gamma$-ray flux from dark matter annihilation is:
\begin{equation}
\label{eq:fluxDM}
\frac{\diff \Phi_{\gamma}}{\diff E}=\frac{1}{8\pi} \frac{\langle \sigma v \rangle}{m_{\text{DM}}^2} \frac{\diff N}{\diff E} \times J(m_{\text{DM}}),
\end{equation} 
where $\diff N/\diff E$ is the spectrum of $\gamma$-rays produced per annihilation event, obtained from PPPC4DM~\cite{Cirelli:2010xx} and dependent on both the dark matter mass and annihilation channel. We constrain the dark matter parameter space by calculating $TS$ values for each combination of cross-section and mass, 
\begin{equation}
\label{eq:tsDM}
    TS \left(\langle \sigma v \rangle, m_{\text{DM}} \right) = -2~{\rm log} \frac{\mathcal{L}_0}{\mathcal{L}\left(\langle \sigma v \rangle, m_{\text{DM}} \right)}.
\end{equation}
We sample 1,000 logarithmically spaced dark matter masses between 10~GeV--10~TeV and 60,000 $\sigmav$ values logarithmically spaced within 5 orders of magnitude around the thermal relic cross-section ($3\times10^{-26}$~cm$^{3}$/s) in both directions. We use J factors from Fig.~\ref{fig:cluster-J} and annihilation channels that produce $b\bar{b}$, $\tau^+\tau^-$, and $W^+W^-$ final states. By stepping through the values of $\langle \sigma v \rangle$, we determine the cross-section value corresponding to a 95\% confidence level (CL) limit, defined as the point where the $TS$ value decreases by 2.71. Detailed analysis results including TS maps, SEDs, and individual cluster results are presented in Appendix~\ref{sec:fermi-results}.



\section{Results}
\label{sec:results}

We find no evidence for a spatially extended $\gamma$-ray signal in any of the seven galaxy clusters listed in Table~\ref{tab:clusters}. We note that, however, when fitting dark matter annihilation spectra to the observed flux limits, several clusters show modest statistical preferences for DM models over background-only scenarios (see Appendix~\ref{sec:fermi-results} for detailed individual cluster analysis). Thus, we place upper limits on the dark matter annihilation cross section using 15 years of \emph{Fermi}-LAT data. 

The Virgo cluster, unsurprisingly, provides the strongest constraints among our targets, primarily due to its large annihilation $J$-factor (Fig.~\ref{fig:cluster-J}). Figure~\ref{fig:comparison} compares our Virgo limits for the $b\bar{b}$ annihilation channel to other relevant benchmarks, showing that we rule out dark matter masses below 
$\sim$200~GeV at a 95\%  CL. Below $\sim$1~TeV, our Virgo limit is stronger by about an order of magnitude compared to the stacking of dwarf spheroidal galaxies, while at higher masses (up to $\sim$10~TeV) it remains competitive \cite{McDaniel:2023bju}.
Finally, our annihilation limits are about an order of magnitude more stringent than previous limits from galaxy clusters \cite{DiMauro:2023qat, Manna:2024dna} and the broader large-scale structure \cite{Bartlett:2022ztj}, which relied on conventional subhalo models that did not account for prompt cusps.


Figure~\ref{fig:ul_clusters} shows our results for annihilation into $b\bar{b}$, $\tau^+\tau^-$ and $W^+W^-$. Besides Virgo, our sample includes Centaurus, NGC~4636, M~49, Fornax (NGC~1339), Hydra, and Coma. For the $b\bar{b}$ channel, our limits exclude the thermal relic cross section for dark matter in the $\sim$20--200~GeV mass range. At higher masses, we observe the sensitivity fall-off characteristic of $\gamma$-ray searches, where the decreasing photon statistics weaken the constraints above a few hundred GeV. Finally, although these clusters have lower $J$-factors (Fig.~\ref{fig:cluster-J}), their smaller spatial extension leads to upper limits that are within an order of magnitude of the Virgo constraints. This strengthens our conclusion by showing that any issue with Virgo (\textit{e.g.}, a complex merger history) does not strongly affect our results.

\section{Conclusion}
\label{sec:concl}

Our results strongly constrain dark matter annihilation in standard cosmological models. The prompt cusps that arise in these scenarios constitute only a small fraction of the dark matter mass, but they dominate the annihilation rate today. We find that, for dark matter masses below a few hundred GeV (annihilating into hadronic final states), our limits from prompt cusps in the Virgo cluster represent the most stringent bounds to date, surpassing those from dSphs and the IGRB. This improvement arises because the annihilation rate in these cusps scales with $\rho$, rather than $\rho^2$, and is therefore enhanced in massive, extended structures such as galaxy clusters. 

These new limits exclude regions of the thermal WIMP parameter space that would remain allowed under conventional substructure assumptions. While typical astrophysical searches focus on centrally dense targets like the Galactic Center or dSphs , our work underscores the crucial role that galaxy clusters can play when 
prompt cusps are taken into account.
Moreover, these constraints, derived under standard cosmological assumptions, not only exceed those from dwarf galaxies but also strongly disfavor the dark matter interpretation of the Galactic Center Excess. 
This comparison is appropriate, as prompt cusps would not boost annihilation to a significant degree in those already-dense regions \cite{Delos:2022bhp,Stucker:2023rjr}.

Furthermore, our results usually provide stronger constraints on dark matter annihilation compared to the IGRB constraints on prompt cusps analyzed in \cite{Delos:2023ipo}. While Delos \textit{et al}.~exclude dark matter masses below $\sim$120~GeV for annihilation into $b\bar{b}$, we extend this exclusion up to $\sim$200~GeV by focusing on galaxy clusters. This improvement is due to our targeting of high $J$-factor regions where the annihilation signal is enhanced compared to the diffuse IGRB.

We note that galaxy clusters can produce $\gamma$-ray emission from cosmic-ray interactions with the intracluster medium, creating potential astrophysical backgrounds for dark matter searches~\cite{2010MNRAS.409..449P, Huber:2013cia, Hussain:2022tls}. While a detailed treatment of these cosmic-ray backgrounds is beyond the scope of this study, our constraints remain robust because prompt cusp annihilation produces distinctive spatial signatures that differ from the extended emission expected from cosmic-ray processes.


Finally, the highly extended, but not entirely isotropic, nature of galaxy cluster searches remains tractable (if challenging) for next-generation atmospheric Cherenkov telescopes, as well as future MeV instruments. Moving forward, extending this analysis to include next-generation $\gamma$-ray observatories with improved sensitivity and angular resolution will be crucial for further probing the dark matter parameter space. These efforts will not only refine our understanding of dark matter, but also strengthen the synergy between observational and theoretical studies in astrophysics and particle physics.







\section*{Acknowledgements}
MC and TL acknowledge support from the Swedish Research Council under contract 2022-04283 and the Swedish National Space Agency under contract 2023-00242. TL also acknowledges sabbatical support from the Wenner-Gren foundation under contract SSh2024-0037.

This project made use of \texttt{Astropy} \cite{astropy:2022}, \texttt{Numpy} \citep{numpy:2020}, \texttt{Matplotlib} \cite{matplotlib:2007}, and \texttt{Pandas} \cite{pandas:2020} Python packages. The authors also acknowledge the use of public data from the \textit{Fermi Science Support Center} data archive. Figure preparation was supported by resources made available by Ciaran O'Hare at \href{https://github.com/cajohare/HowToMakeAPlot}{\texttt{https://github.com/cajohare/HowToMakeAPlot}} (last accessed on January 20, 2025). The color scheme used in our plots was inspired by \emph{Cooper \& Gorfer - ``The Garden,''} exhibited at the Fotografiska Museum in Stockholm in 2021.

\appendix
\renewcommand\thesection{\Alph{section}}
\section{Pericenter distribution}
\label{sec:peri}

In Sec.~\ref{sec:signal}, we assumed that cusp orbital pericenters are uniformly distributed in radius.
Here we evaluate the pericenter distribution for idealized halos and show that the pericenter radius distribution is indeed close to uniform.
For this purpose, consider a halo with an isotropic distribution function $\df(E)$.
That implies that at any fixed radius $r$, the velocity distribution is
\begin{equation}
    \diff P = \frac{\df(\Phi(r)+\vec v^2/2)}{\rho(r)}\diff^3\vec v,
\end{equation}
where $\Phi(r)$ is the gravitational potential and $\rho(r)$ is the density profile. We write this distribution as a differential probability $\diff P$, a useful representation as we transform the distribution to other variables.

Now consider the velocity in spherical coordinates aligned along the radius vector, taking $\theta$ to be the angle between $\vec v$ and $\vec r$, so that $\diff^3\vec v=2\pi\,\diff\cos\theta\, v^2\diff v$. Here we integrate out the azimuthal angle, since it is irrelevant to the calculation. The polar angle $\theta$ is relevant, however, because by conservation of energy and angular momentum, it is related to the pericenter radius $\rp$ through
\begin{equation}\label{rp}
    \Phi(r)-\Phi(\rp) -\left(\sin^2\theta\frac{r^2}{\rp^2}- 1\right)\frac{v^2}{2} = 0.
\end{equation}
By inserting a Dirac delta function $\ddelta$ to enforce Eq.~(\ref{rp}) and appropriately scaling by the derivative of Eq.~(\ref{rp}) with respect to $\rp$, we can expand the differential probability as
\begin{align}
    \diff P &= \frac{\df(\Phi(r)+v^2/2)}{\rho(r)}2\pi\,\diff\cos\theta\, v^2\diff v
    \nonumber\\&\hphantom{=}\times
    \ddelta\!\left\{\Phi(r)-\Phi(\rp) -\left(\sin^2\theta\frac{r^2}{\rp^2}- 1\right)\frac{v^2}{2}\right\}
    \nonumber\\&\hphantom{=}\times
    \left|\sin^2\theta\frac{r^2}{\rp^3}v^2-\Phi^\prime(\rp)\right|\diff \rp,
\end{align}
where $\Phi^\prime$ is the derivative of the potential.
To obtain the pericenter distribution, all that remains is to integrate out $v$ and $\theta$.
The $\theta$ integral yields
\begin{align}
    \diff P &= \frac{\df(\Phi(r)+v^2/2)}{\rho(r)}4\pi v^2\diff v\,
    \nonumber\\&\hphantom{=}\times
    \step\!\left[v^2-2\rp^2\frac{\Phi(r)-\Phi(\rp)}{r^2-\rp^2}\right]
    \nonumber\\&\hphantom{=}\times
    \frac{\rp}{rv}
    \frac{|v^2+2(\Phi(r)-\Phi(\rp))-\rp\Phi^\prime(\rp)|}{\sqrt{(r^2-\rp^2)v^2-2\rp^2(\Phi(r)-\Phi(\rp))}}
    \diff \rp,
\end{align}
where the Heaviside step function $\step$ selects the velocities $v$ for which Eq.~(\ref{rp}) has solutions.
The distribution of pericenter radii $\rp$ is thus
\begin{align}
    \frac{\diff P}{\diff\rp} &=
    \frac{4\pi\, \rp}{\rho(r)r}
    \int_{v_\mathrm{min}(\rp,r)}^\infty
    \diff v\,v
    \df(\Phi(r)+v^2/2)
    \nonumber\\&\hphantom{=}\times
    \frac{|v^2+2(\Phi(r)-\Phi(\rp))-\rp\Phi^\prime(\rp)|}{\sqrt{(r^2-\rp^2)v^2-2\rp^2(\Phi(r)-\Phi(\rp))}},
\end{align}
where $v_\mathrm{min}(\rp,r)\equiv\sqrt{2\rp^2[\Phi(r)-\Phi(\rp)]/(r^2-\rp^2)}$.

\begin{figure}[tbp]
\centering
\includegraphics[width=\columnwidth]{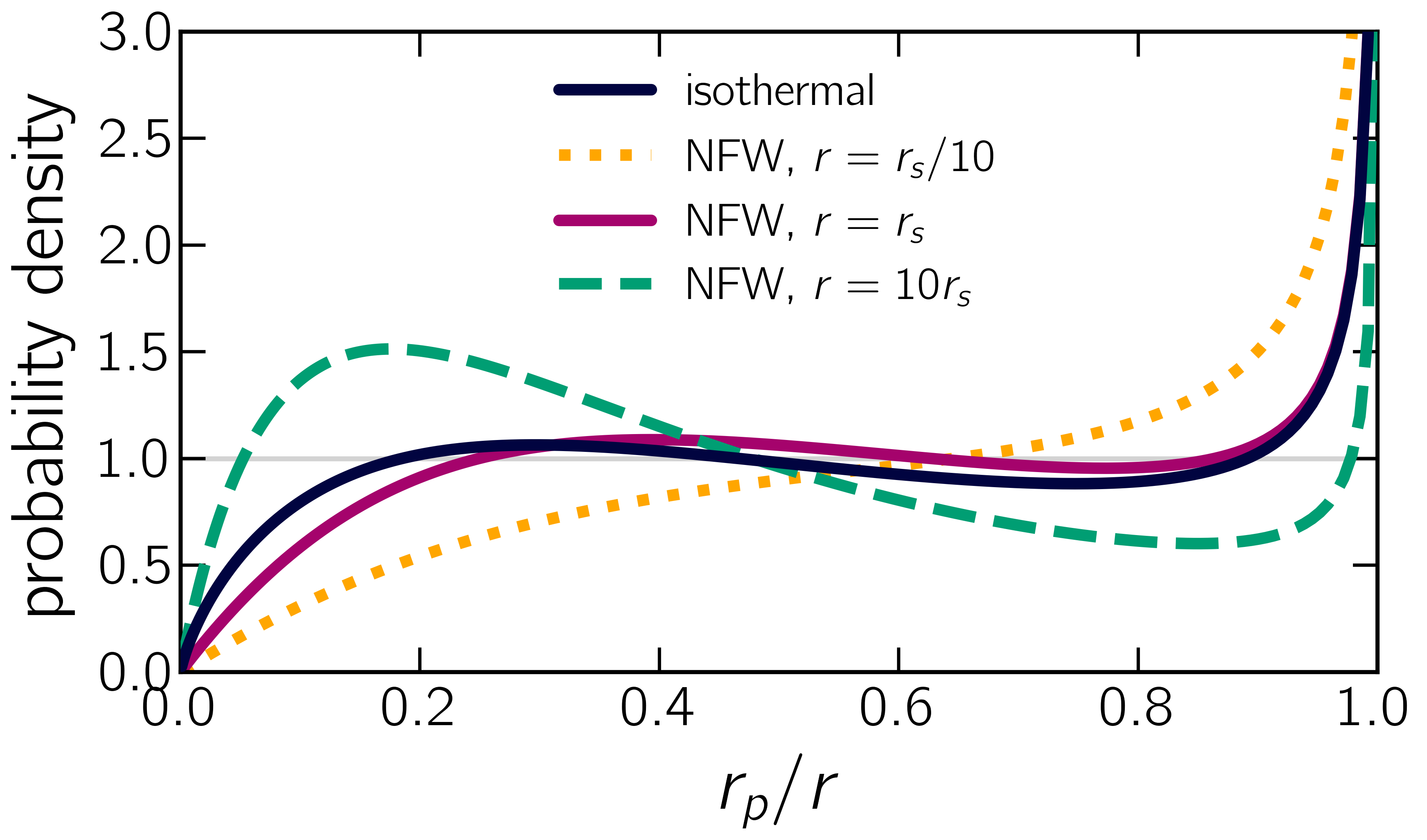}
\caption{Distributions of pericenter radii for particles in idealized halos. We consider an isothermal sphere (black curve) and particles at several set radii inside NFW halos with isotropic velocity distributions (colored curves). For reference, the thin horizontal line indicates a uniform distribution. The pericenter distributions tend to be close to uniform in radius.}
\label{fig:peri}
\end{figure}

For a singular isothermal sphere, with distribution function $f(E)=\e^{-E/\sigma^2}/[(2\pi)^{5/2}G\sigma]$, potential $\Phi(r)=2\sigma^2\ln(r)$, and density profile $\rho(r)=\sigma^2/(2\pi G r^2)$ (for some set velocity dispersion $\sigma$), the pericenter distribution can be evaluated analytically, yielding
\begin{equation}
    \frac{\diff P}{\diff\rp} =
    \frac{[4 \ln(r/\rp)-1]r^2 +\rp^2}{(r^2-\rp^2)^{3/2}}
    \left(\frac{\rp}{r}\right)^{\frac{r^2+\rp^2}{r^2-\rp^2}}.
\end{equation}
We show this distribution as the black curve in Fig.~\ref{fig:peri}.
For a more realistic NFW density profile, we evaluate the pericenter distribution numerically using the distribution function given as a fitting function by Ref.~\cite{2000ApJS..131...39W}. For several radii $r$, we show the resulting pericenter distributions as the colored curves in Fig.~\ref{fig:peri}.

\section{\textit{Fermi}-LAT Analysis}
\label{sec:fermi-results}
Here, we provide supplementary details from our {\tt fermipy}-based $\gamma$-ray analysis of the seven galaxy clusters listed in Table~\ref{tab:clusters}. The analysis steps are summarized below, followed by individual cluster results.

We analyze 15 years of \texttt{Pass8} \textit{Fermi}-LAT data (MET$_\text{start}=239557417$ to MET$_\text{end}=720502740$), covering energies from 500~MeV to 500~GeV. The data are processed using the \texttt{fermipy} software package (v1.2.0) built on \texttt{FermiTools} (v2.2.0). The main analysis steps are as follows:
\begin{itemize}
\item \textbf{Data Selection}: We select photon events from the \texttt{P8R3\_SOURCE\_V3} event class.
\item \textbf{Quality Cuts}: We apply standard quality cuts, including a zenith angle cut of 90$^\circ$ to mitigate contamination from the Earth's limb.
\item \textbf{Binning}: We bin the data spatially with a pixel size of 0.2$^\circ$, and in energy with 8 logarithmically spaced bins per decade.
\item \textbf{RoI}: We define a 20$^\circ\times20^\circ$ RoI centered on each cluster's dynamical center listed in Table~\ref{tab:clusters}.
\item \textbf{Source Modeling}: All confirmed point and extended sources were included in our background model using the \texttt{4FGL-DR4} parameters \cite{Fermi-LAT:2022byn}. We model sources within a 25$^\circ$ radius of the RoI center to account for contributions from the tails of the \textit{Fermi}-LAT point spread function.
\item \textbf{Diffuse Emission Models}: We incorporate the standard Galactic diffuse emission model (\texttt{gll\_iem\_v07.fits}) and the isotropic diffuse background model appropriate for the event class (\texttt{iso\_P8R3\_SOURCE\_V3\_v1.txt}).
\end{itemize}

We model each galaxy cluster as an extended source using a spatial template derived from the prompt cusp annihilation profiles (Section~\ref{sec:signal}). This component is added to the model as a \texttt{SpatialMap}, with a fixed spectral index of 2.0 and a free normalization parameter.

After building the sky model, we optimize source parameters by freeing the normalizations (\texttt{norm}) of all catalog sources within $5^\circ$ of the RoI center and with TS~$>$25,
as well as all spectral parameters of catalog sources within $7^\circ$ of the RoI center and with TS~$>$~500. The normalization parameters of the diffuse components (Galactic and isotropic) are also left free to vary. To reduce computation time and prevent instability in the fitting procedure, we disable energy dispersion corrections for these diffuse components. This is justified because energy dispersion effects for smooth, extended backgrounds typically result in negligible improvement in fit quality. For the cluster templates and point sources within the analysis region, energy dispersion remains enabled.

We then perform a maximum likelihood fit using \texttt{gta.fit()}, which adjusts all the free parameters to best match the data. We extract signal or spectral upper limits using the SED likelihood procedure, \texttt{gta.sed()}.

\subsection{Dark Matter Analysis}

We convert the measured $\gamma$-ray flux limits to dark matter constraints using the expression in Eq.~(\ref{eq:fluxDM}) and then, for each combination of dark matter mass and velocity-weighted annihilation cross-section, we calculate the TS in Eq.~(\ref{eq:tsDM}). We consider three annihilation channels ($b\bar{b}$, $\tau^+\tau^-$, and $W^+W^-$) using $\gamma$-ray spectra from \texttt{PPPC4DM}~\cite{Cirelli:2010xx} with electroweak corrections enabled.

\begin{figure*}[htbp]
\centering
\begin{tabular}{ccc}
\textbf{TS Map} & \textbf{DM Constraints} & \textbf{SED} \\[0.2cm]
\includegraphics[width=0.29\textwidth]{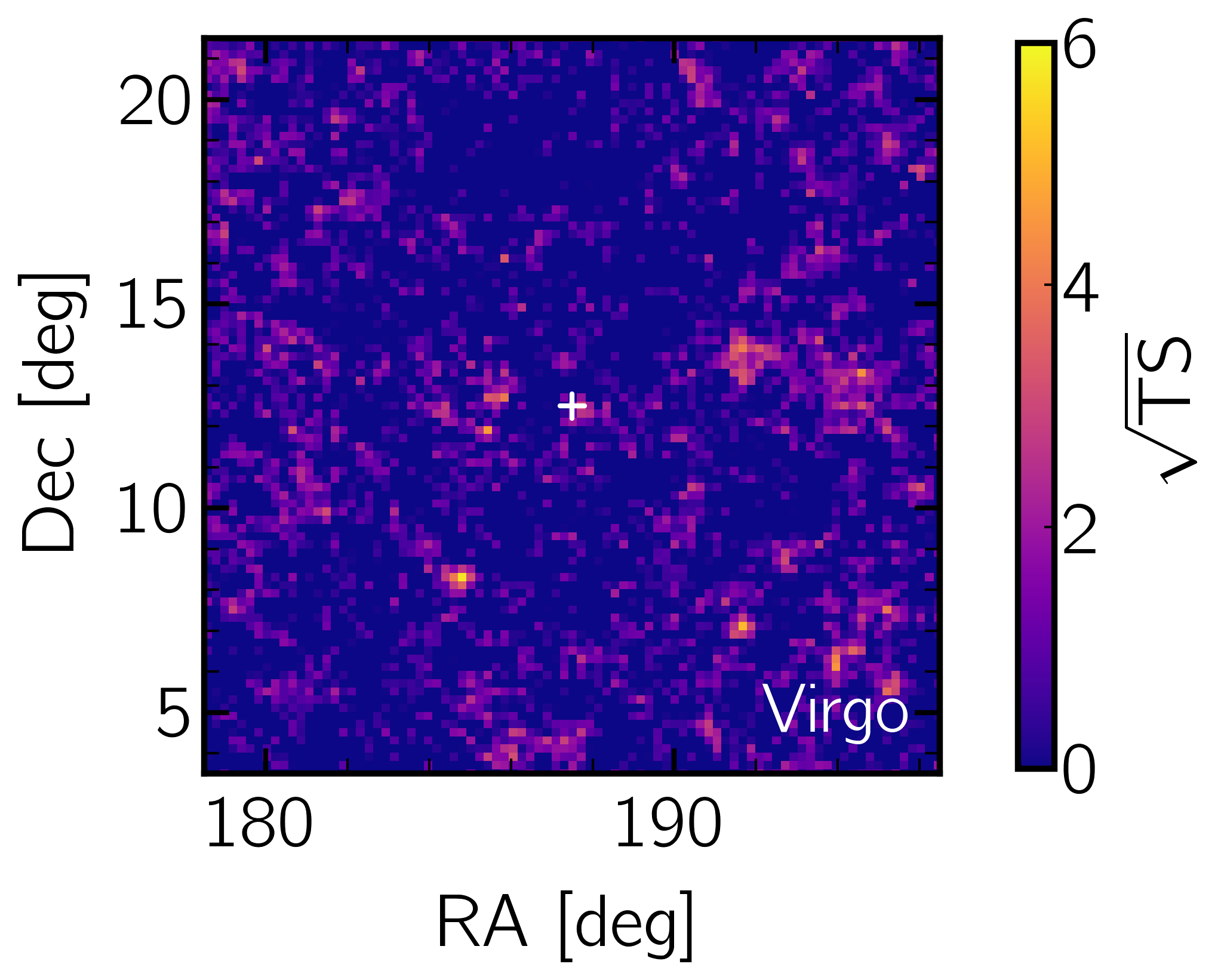} &
\includegraphics[width=0.32\textwidth]{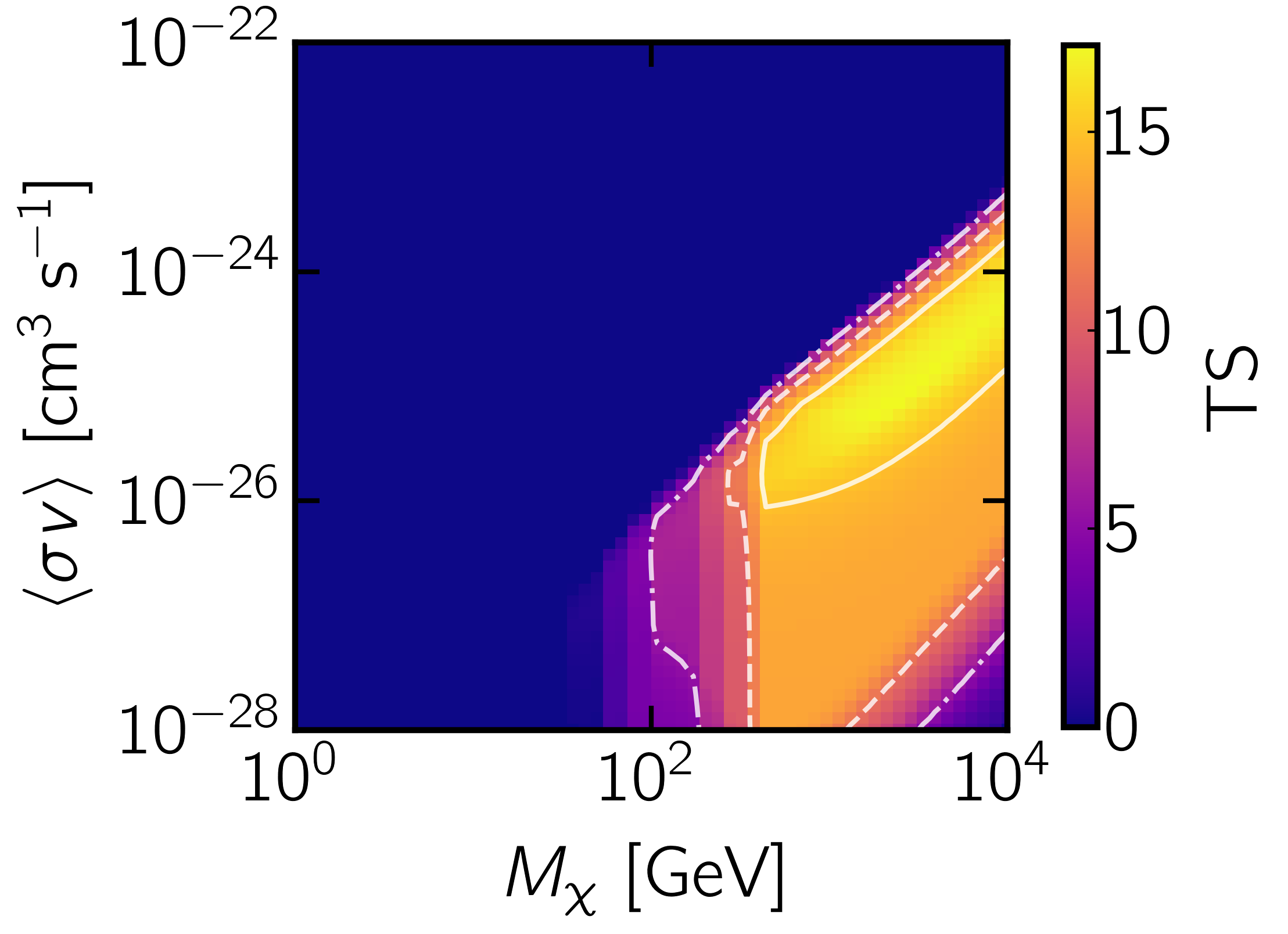} &
\includegraphics[width=0.32\textwidth]{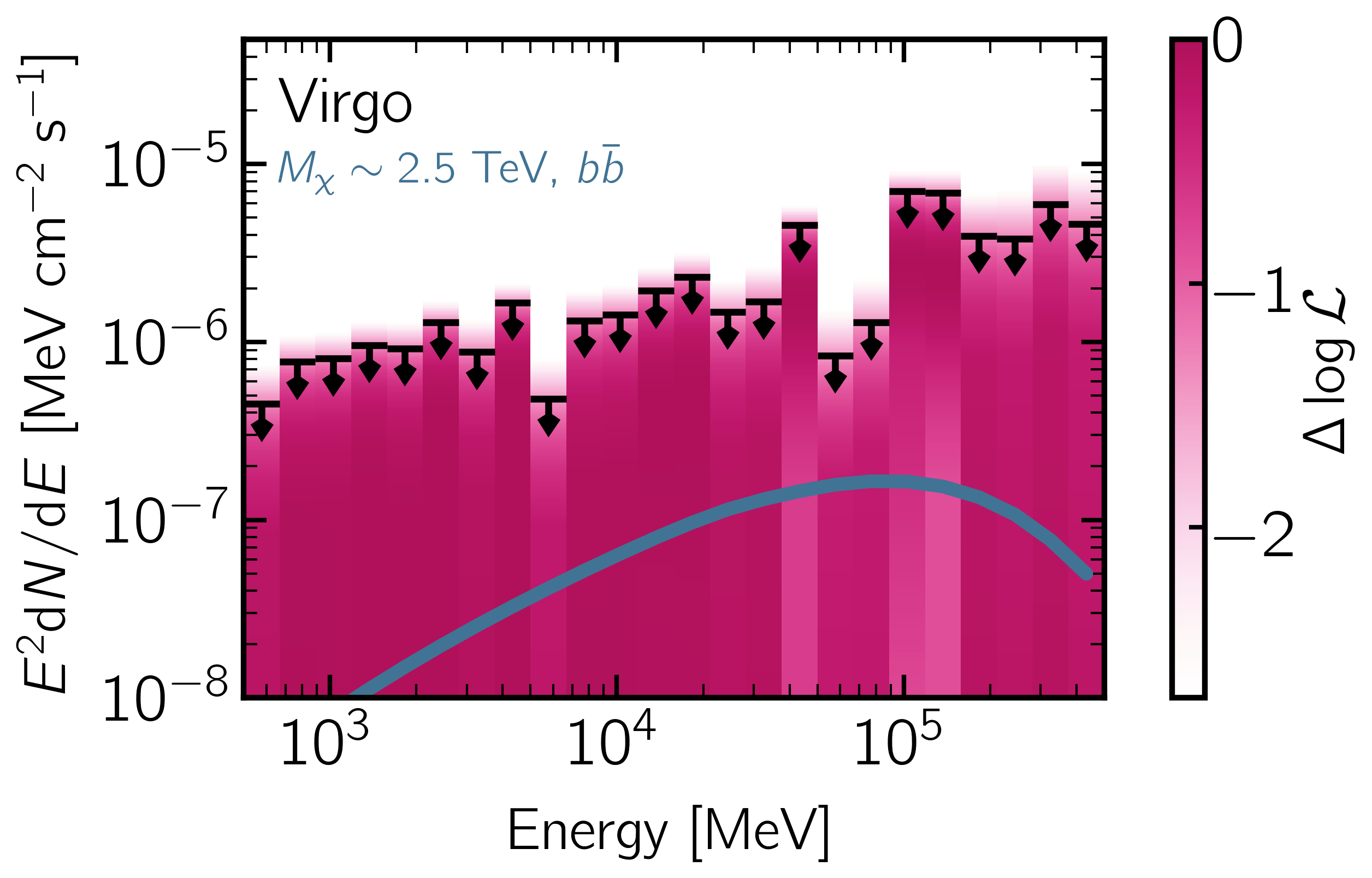} \\[0.3cm]
\includegraphics[width=0.31\textwidth]{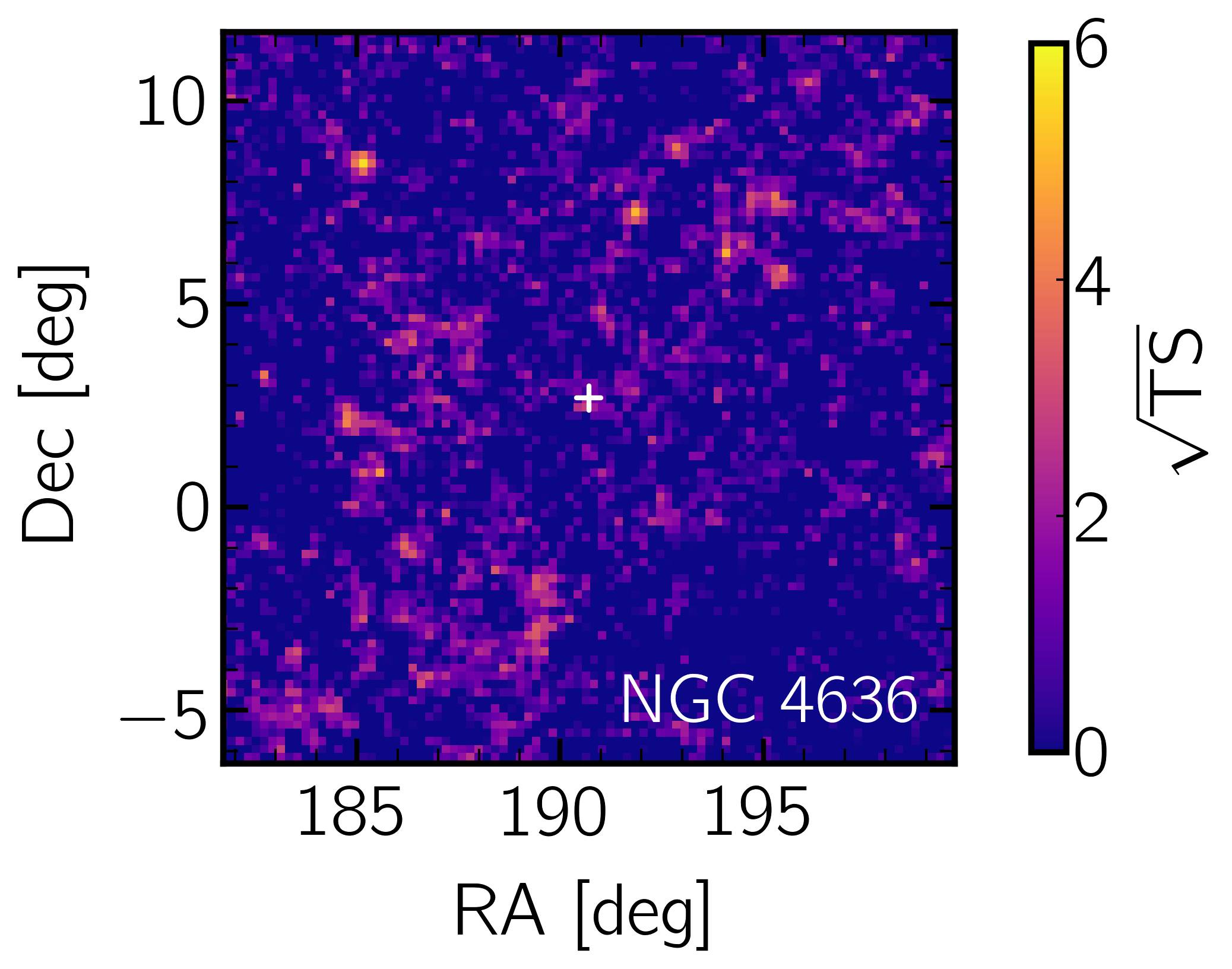} &
\includegraphics[width=0.31\textwidth]{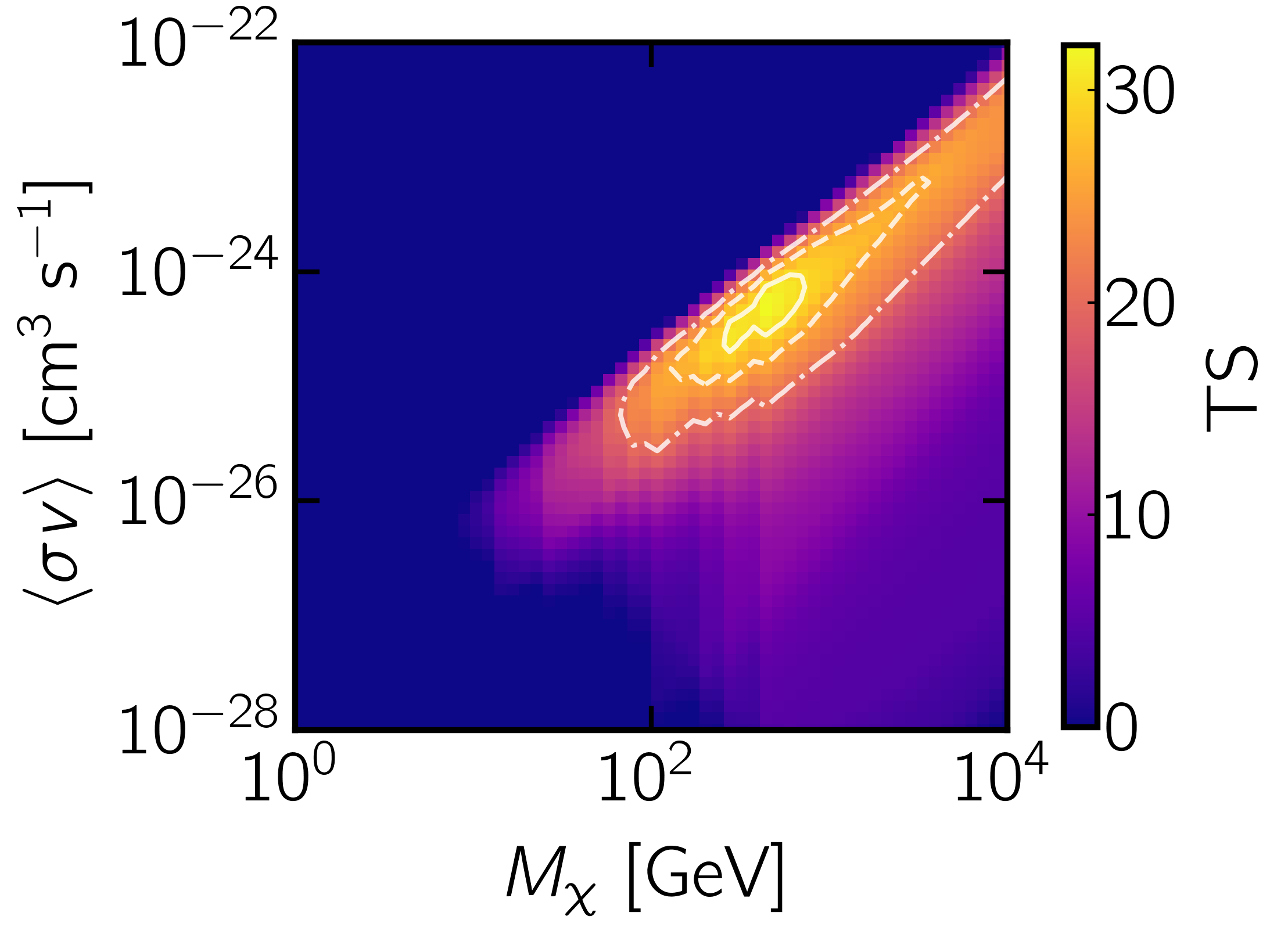} &
\includegraphics[width=0.31\textwidth]{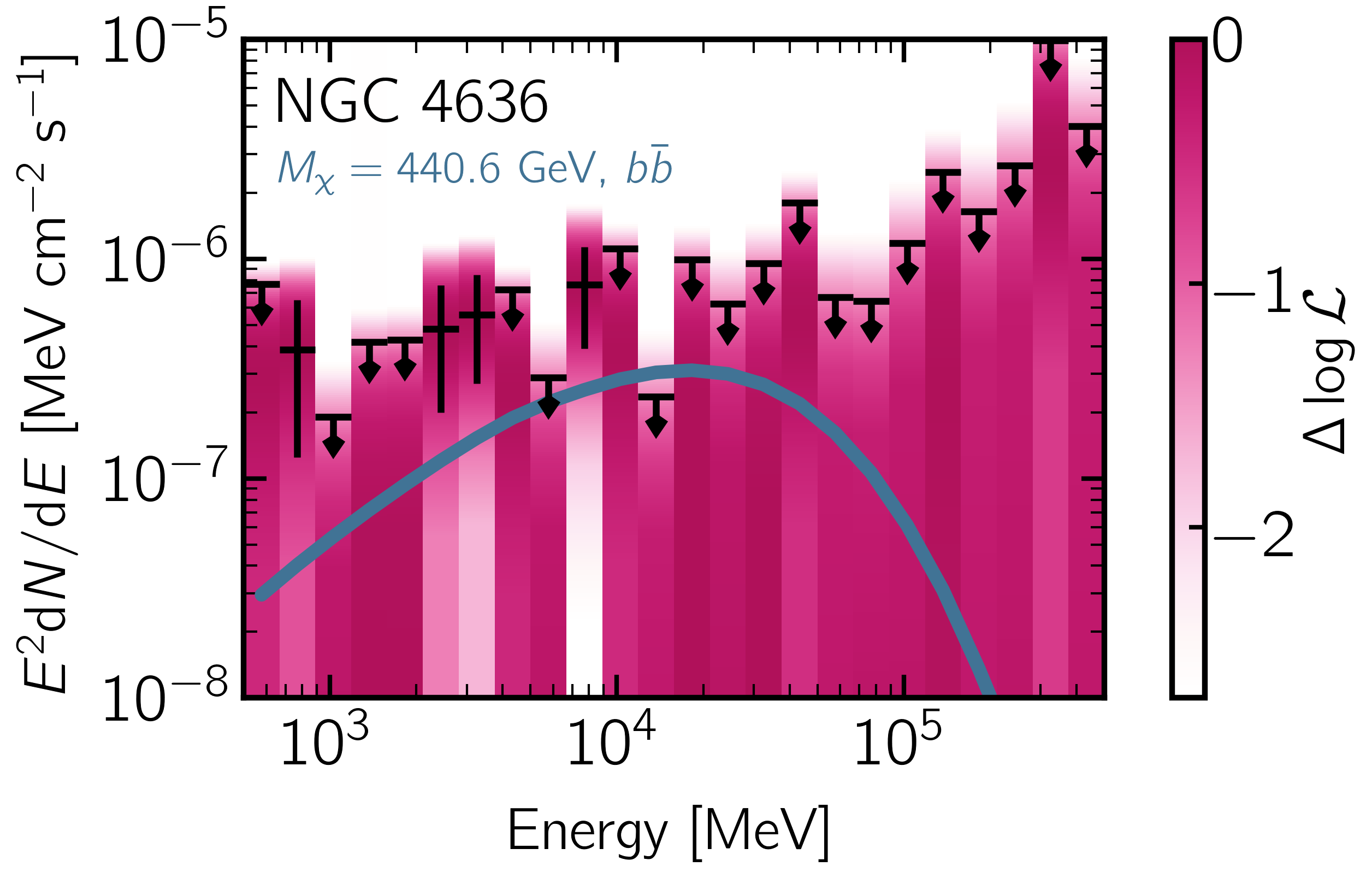} \\[0.3cm]
\includegraphics[width=0.29\textwidth]{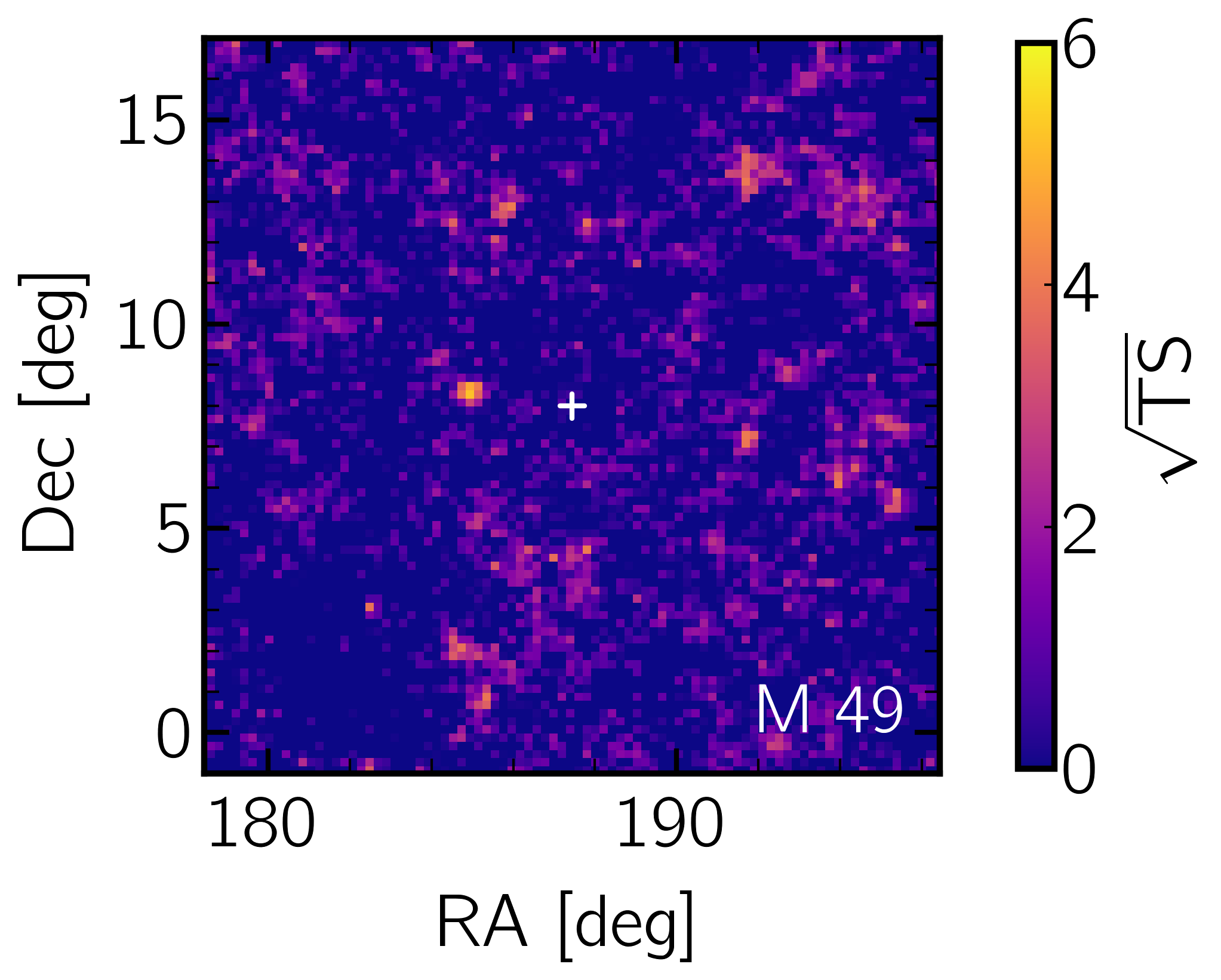} &
\includegraphics[width=0.32\textwidth]{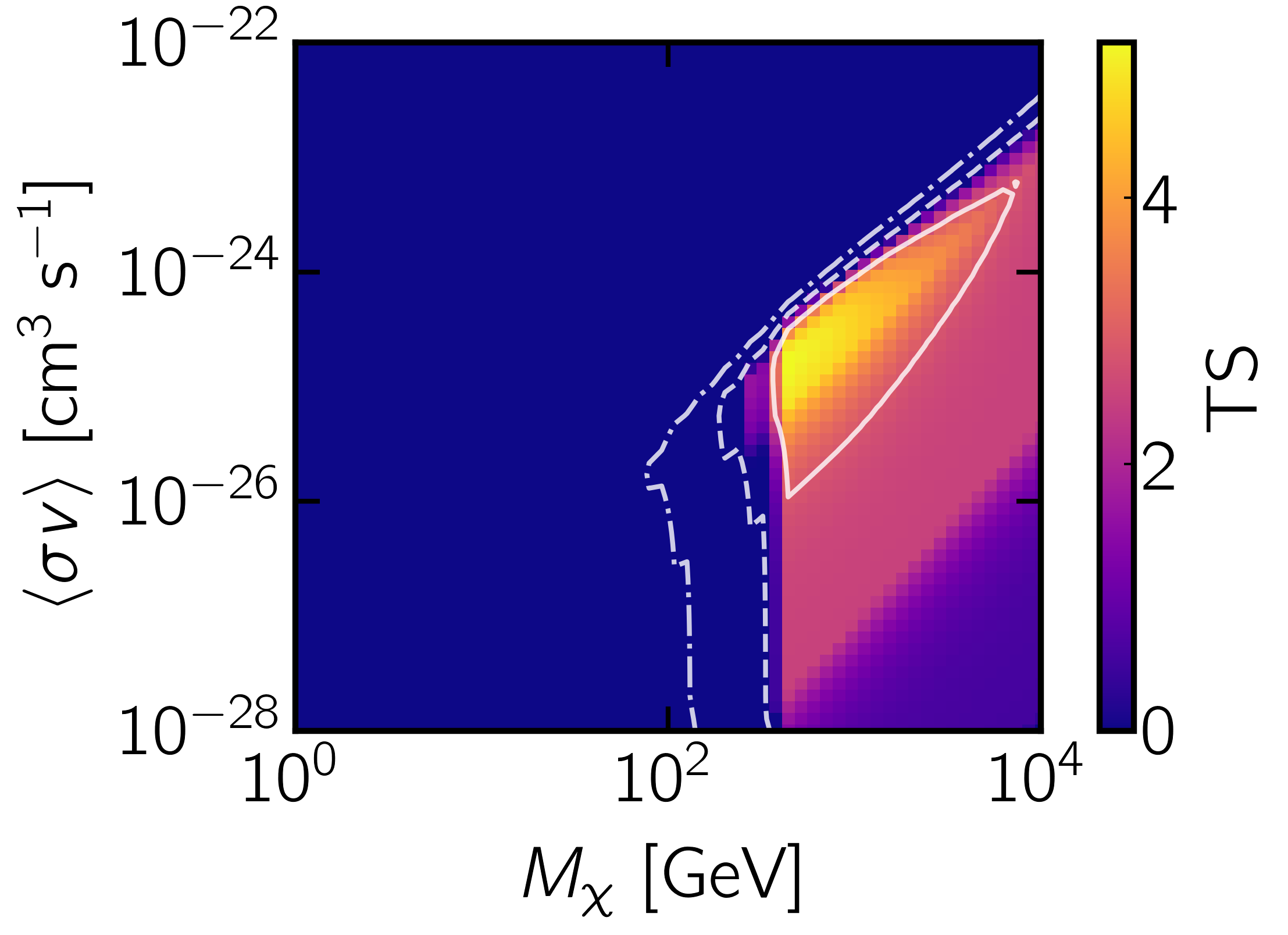} &
\includegraphics[width=0.32\textwidth]{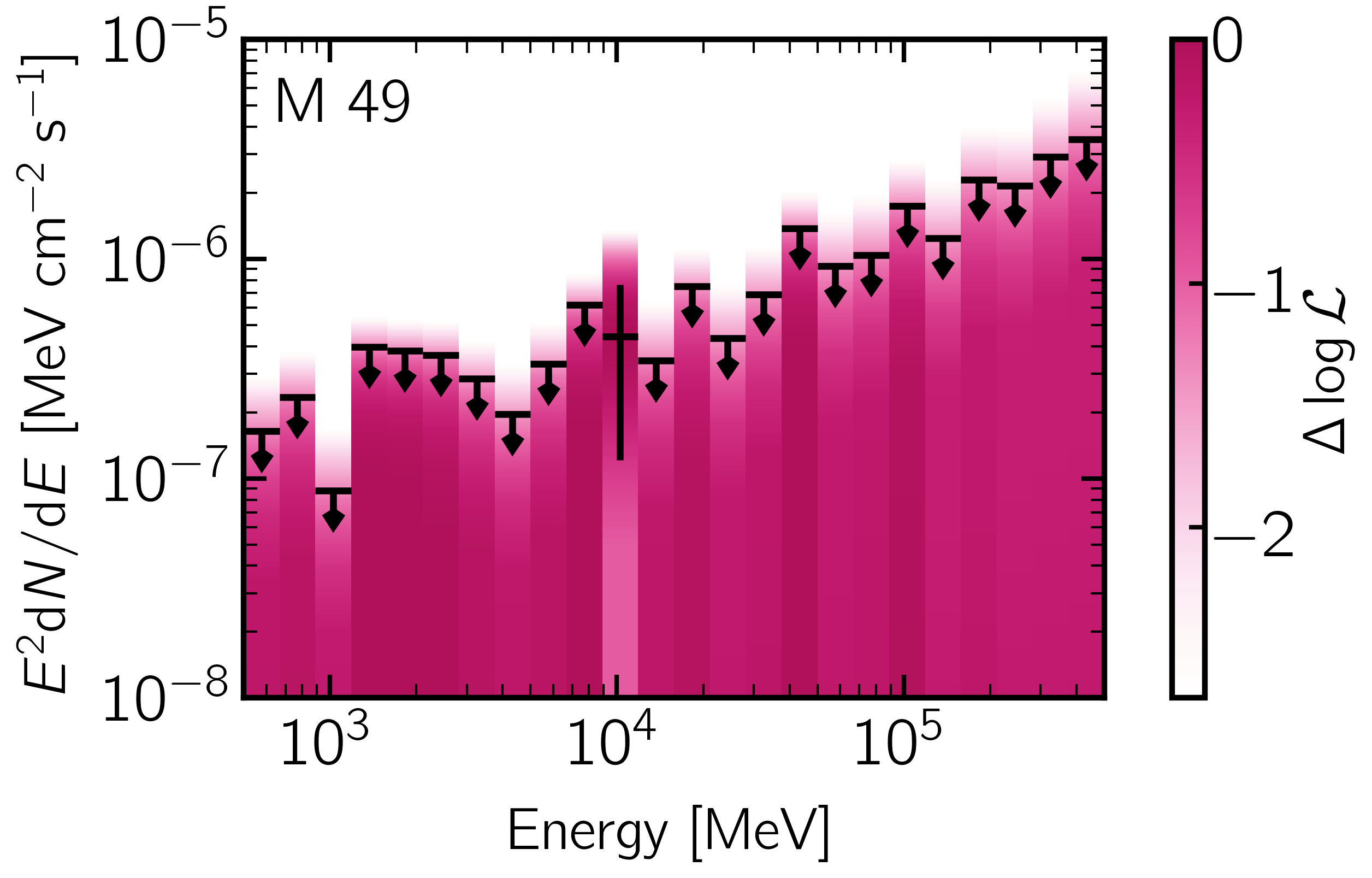} \\[0.3cm]
\includegraphics[width=0.29\textwidth]{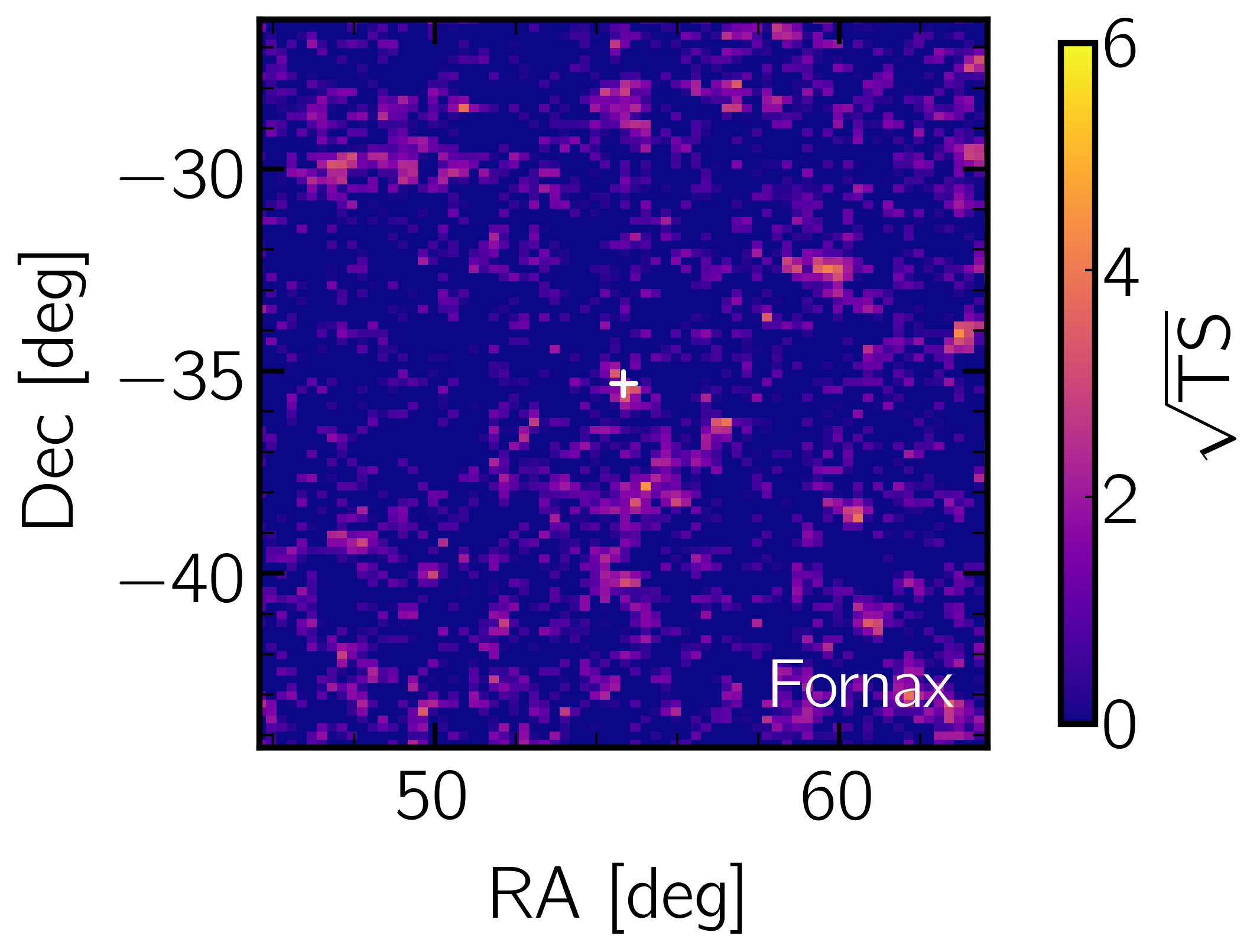} &
\includegraphics[width=0.32\textwidth]{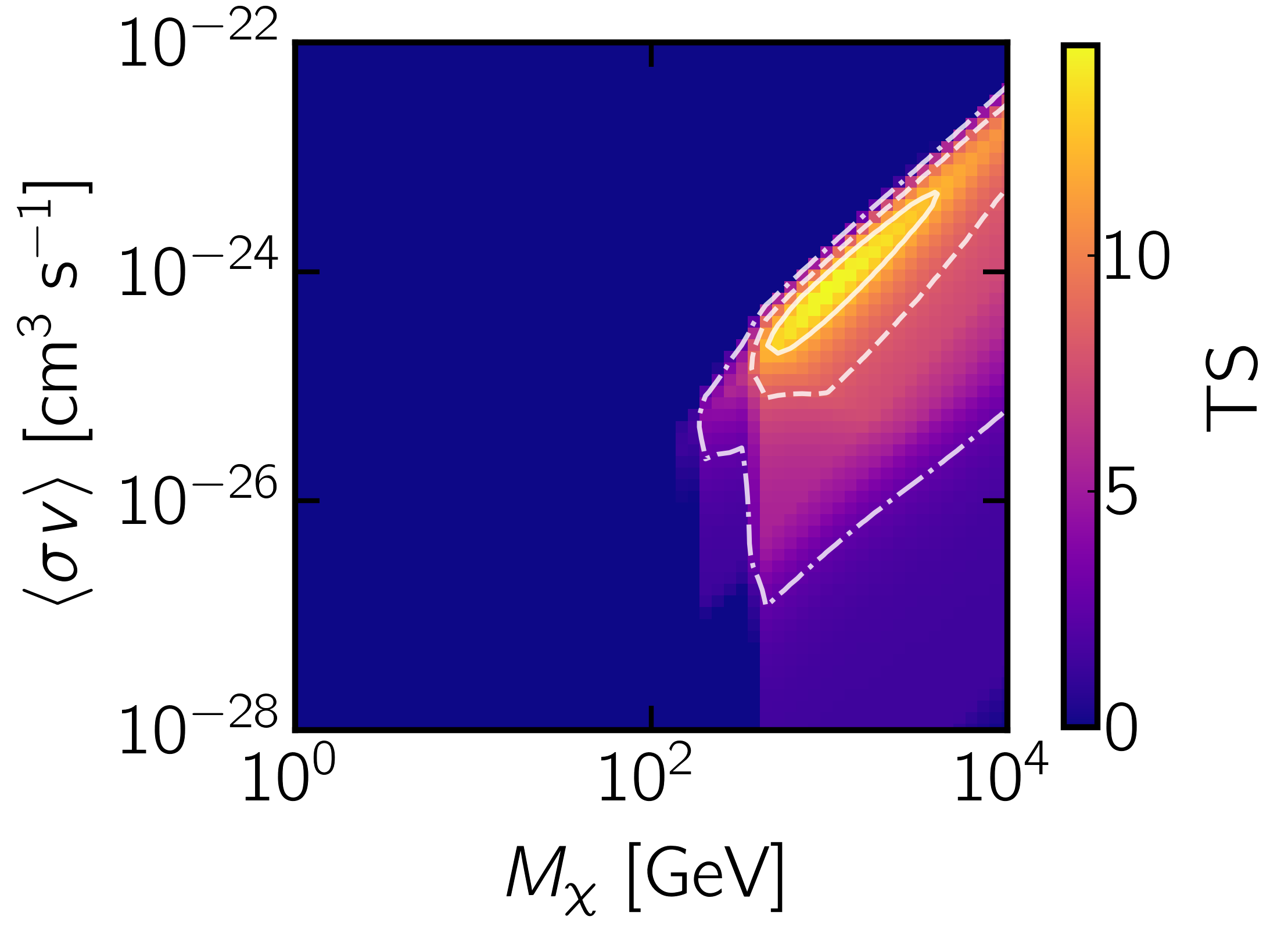} &
\includegraphics[width=0.32\textwidth]{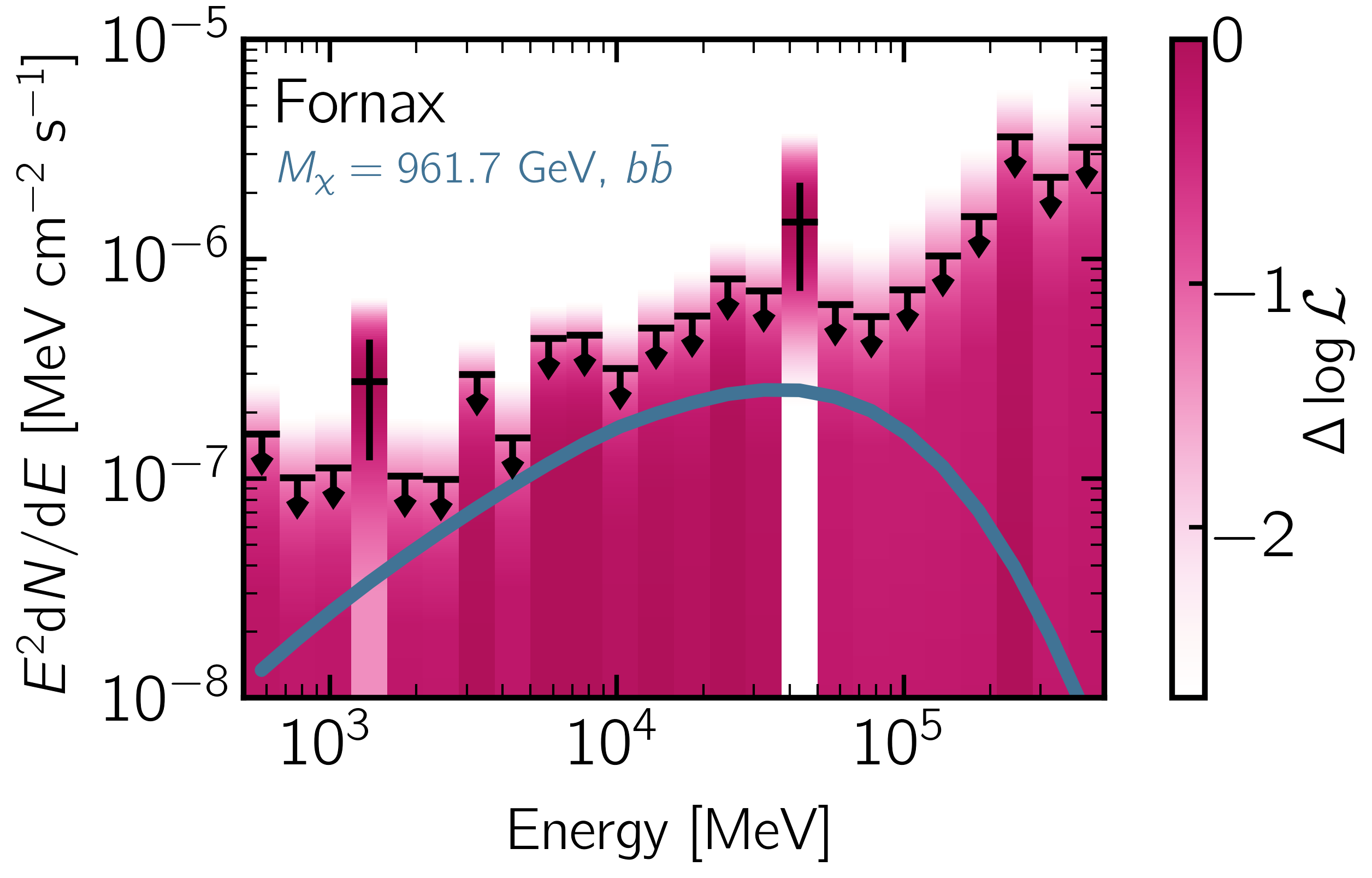} \\[0.3cm]

\end{tabular}
\caption{Comprehensive analysis results for all seven galaxy clusters. \textbf{Left column:} Test statistic ($\sqrt{\text{TS}}$) maps showing no significant $\gamma$-ray excesses in any cluster region. The white cross marks each cluster's dynamical center. \textbf{Middle column:} Dark matter annihilation cross-section constraints as a function of mass for the $b\bar{b}$ channel, with contours showing 68\%, 95\%, and 99.7\% confidence levels. White solid contours indicate the strongest exclusion regions. \textbf{Right column:} Spectral energy distributions showing flux upper limits (black arrows) and likelihood profiles (colored bands) across energy bins from 500~MeV to 500~GeV. The blue curve in Virgo, Fornax, and Coma panels shows the expected spectrum for the best-fit dark matter model.}
\label{fig:lat_results}
\end{figure*}

\renewcommand\thefigure{7 continued}

\begin{figure*}
\centering
\begin{tabular}{ccc}
\textbf{TS Map} & \textbf{DM Constraints} & \textbf{SED} \\[0.2cm]
\includegraphics[width=0.29\textwidth]{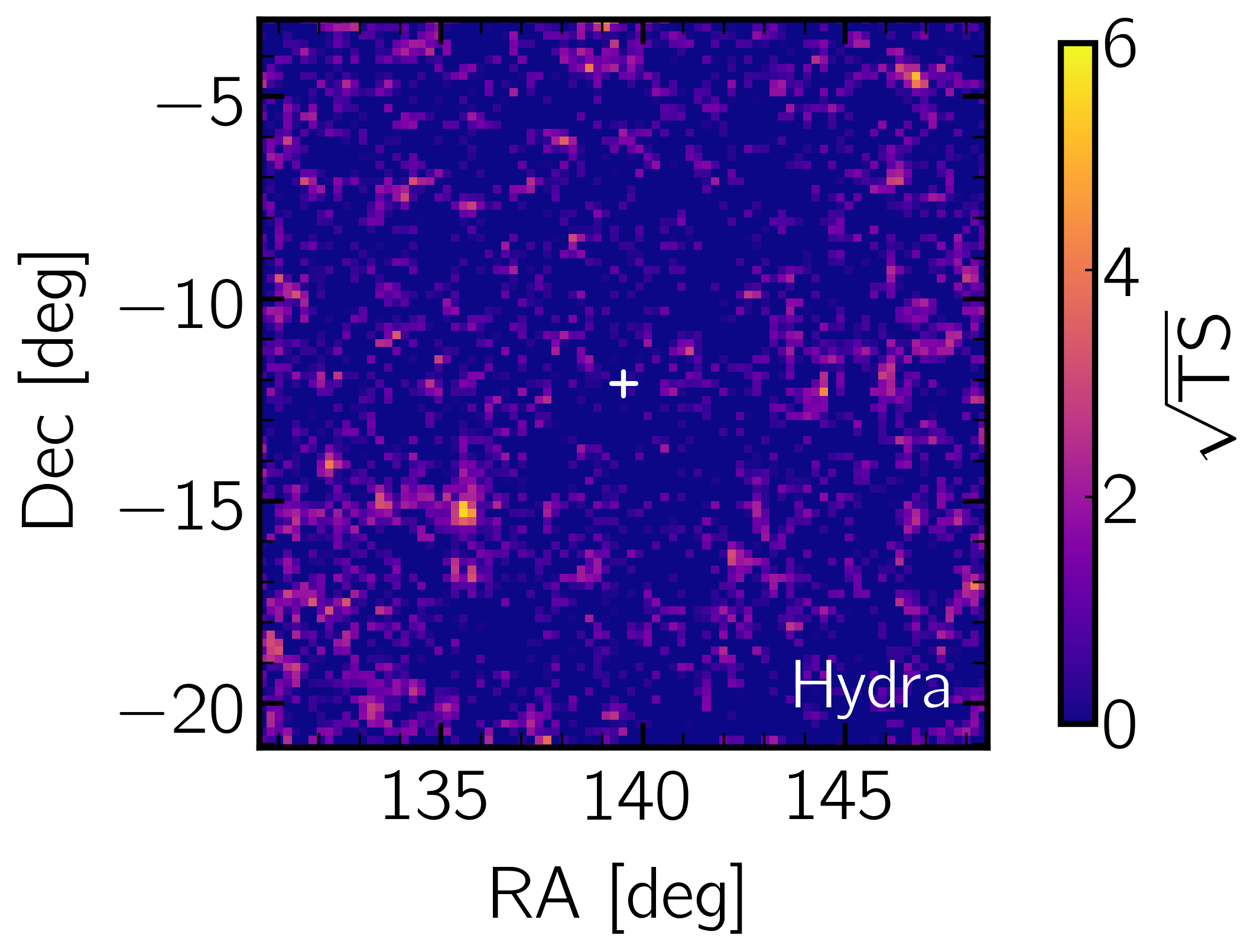} &
\includegraphics[width=0.32\textwidth]{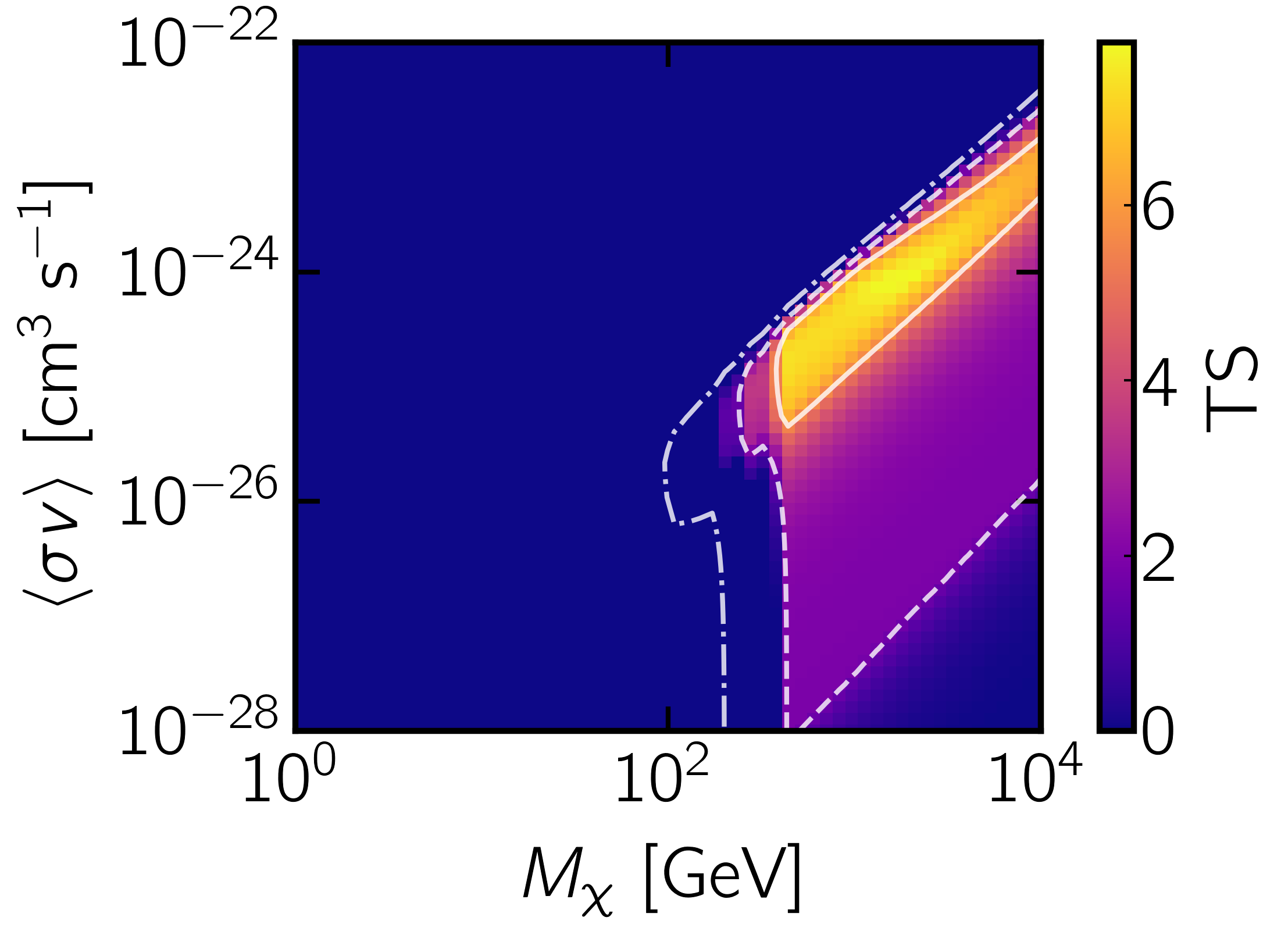} &
\includegraphics[width=0.32\textwidth]{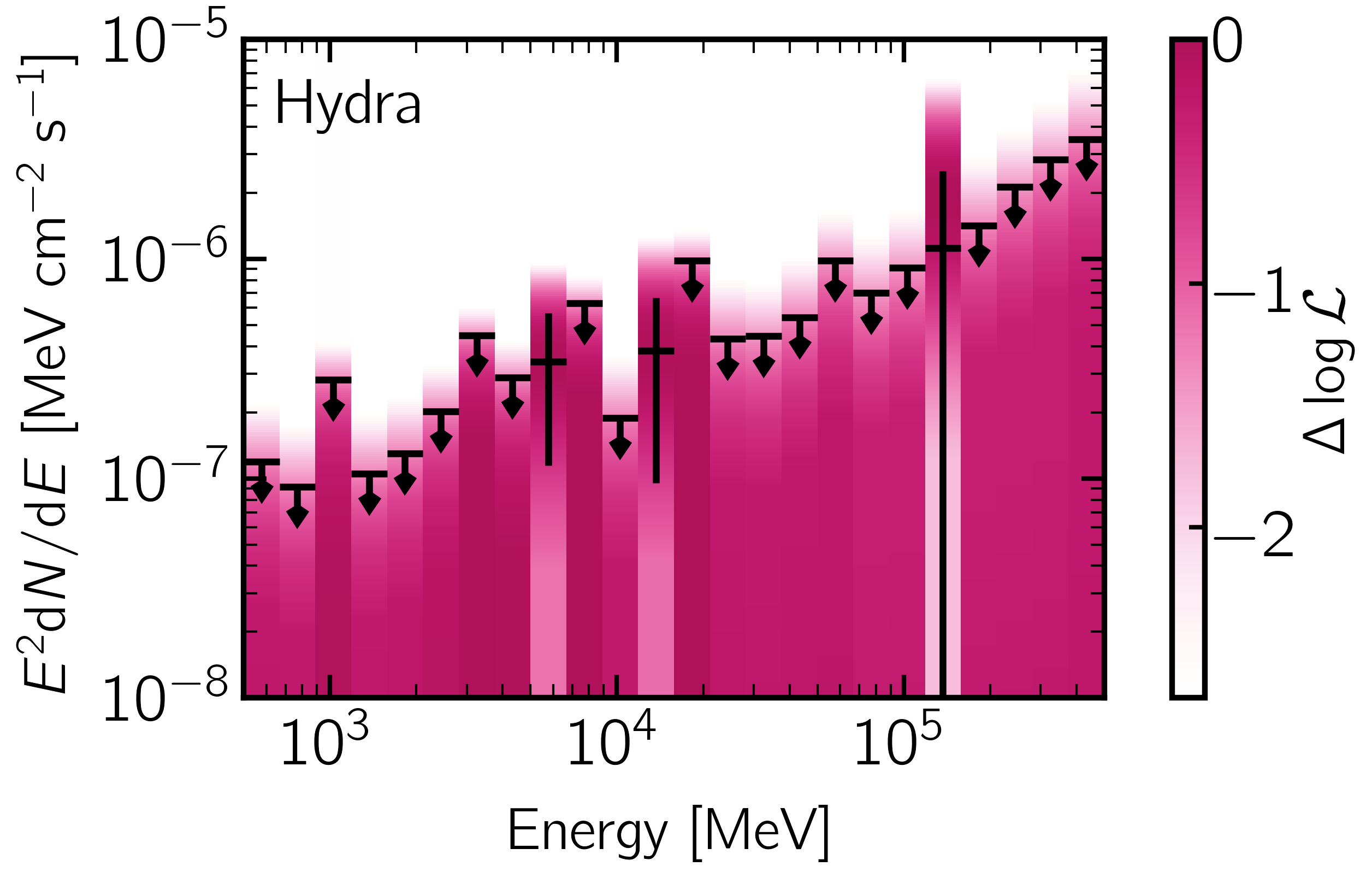} \\[0.3cm]
\includegraphics[width=0.29\textwidth]{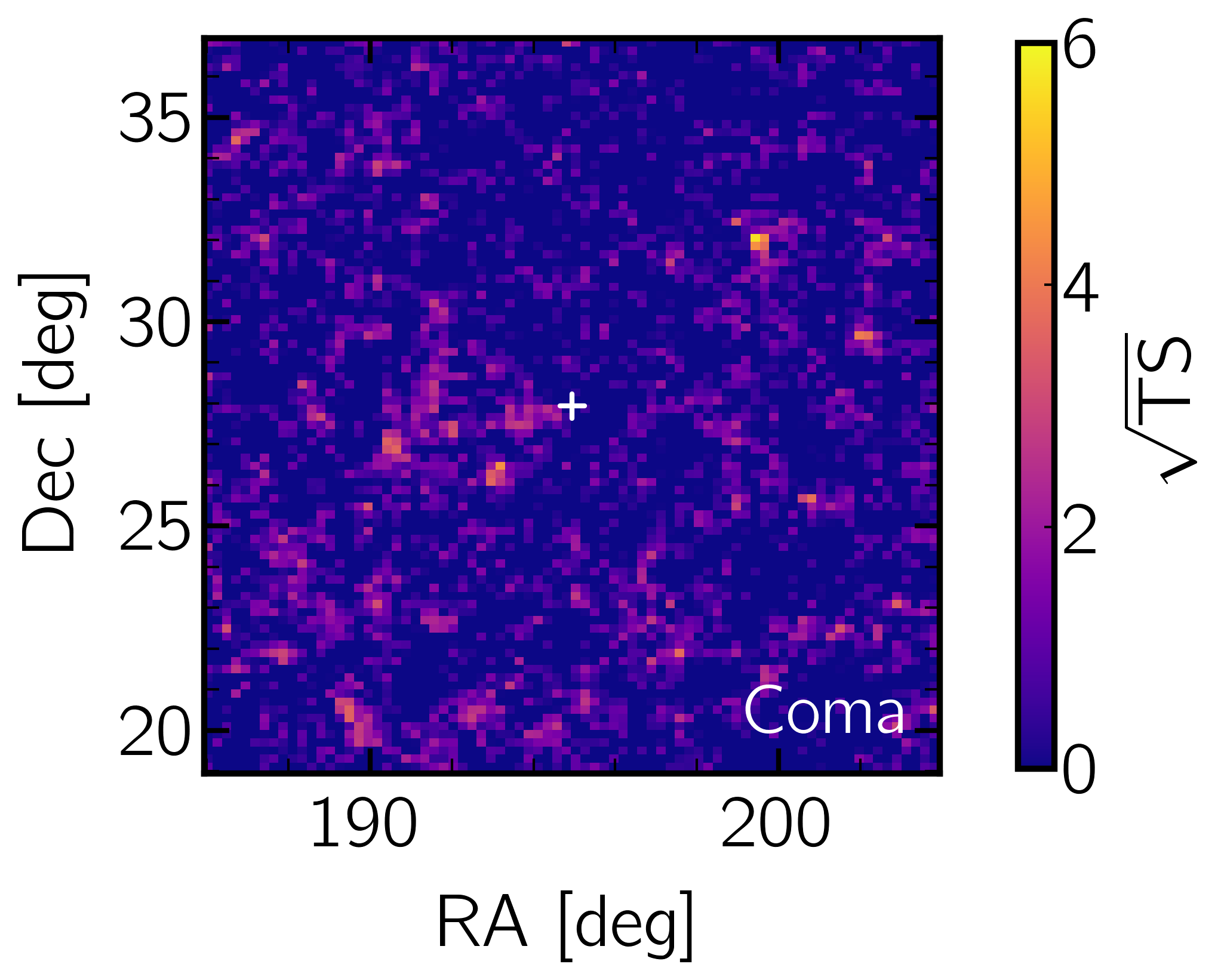} &
\includegraphics[width=0.32\textwidth]{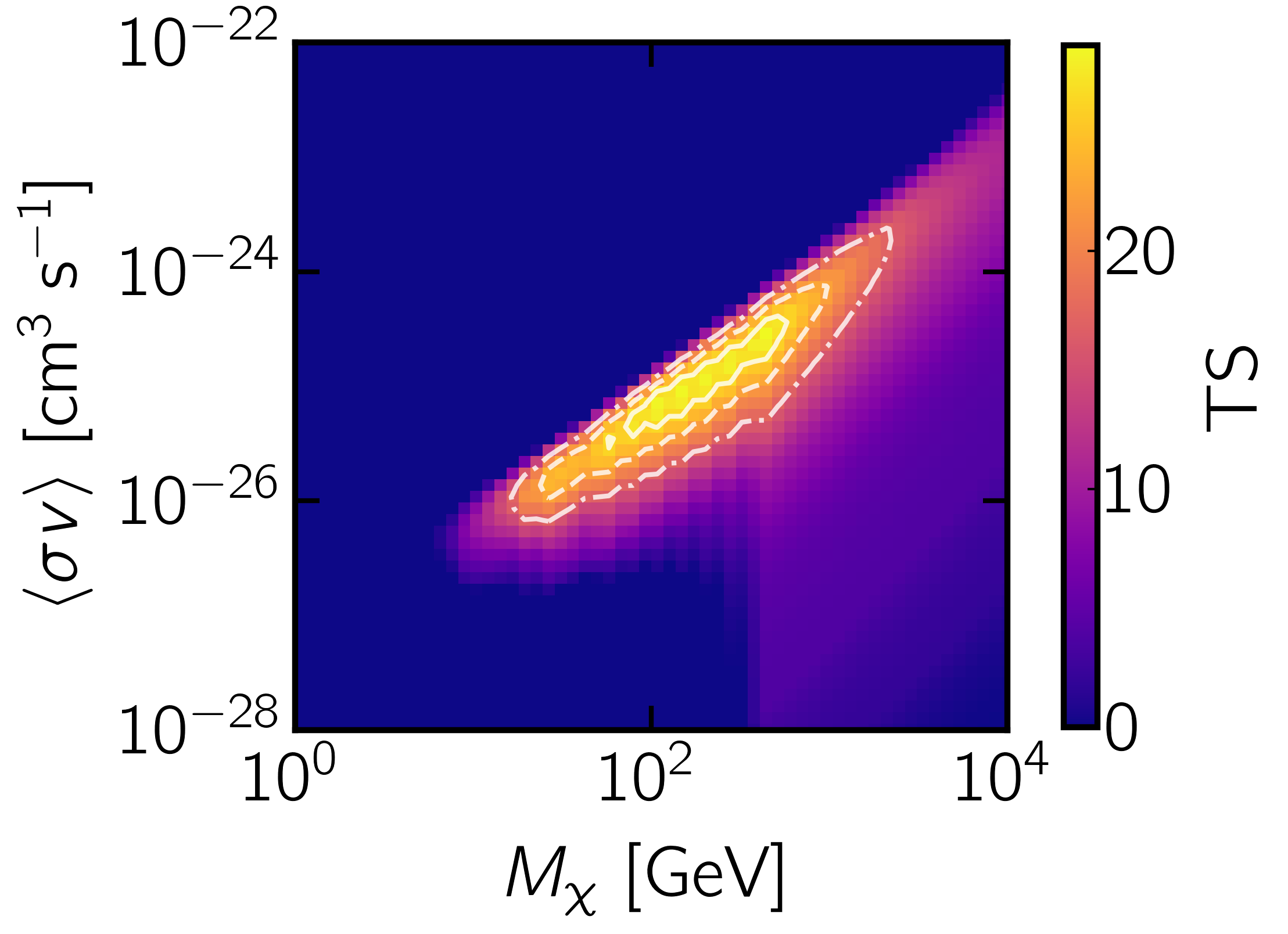} &
\includegraphics[width=0.32\textwidth]{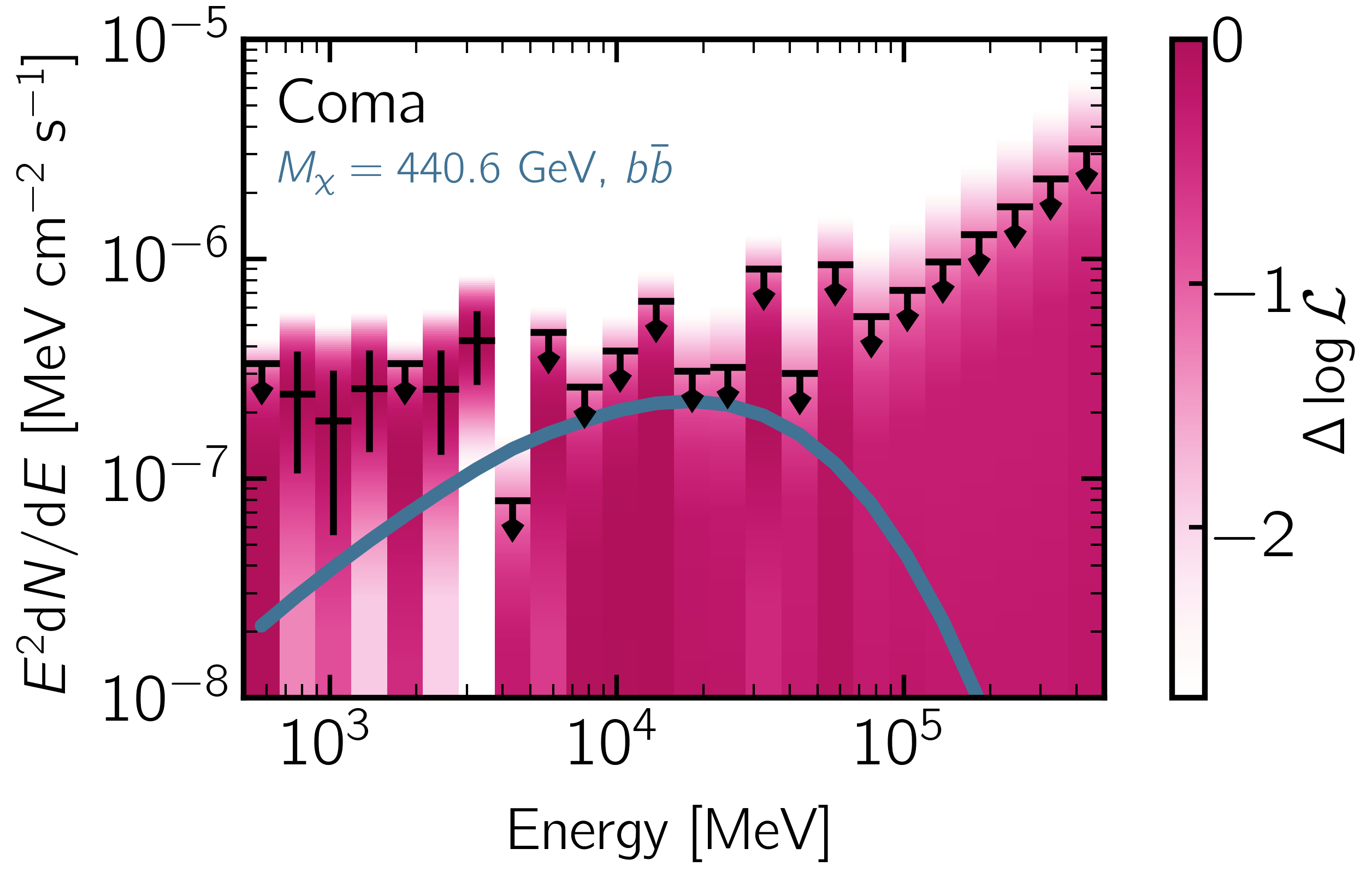} \\
\includegraphics[width=0.29\textwidth]{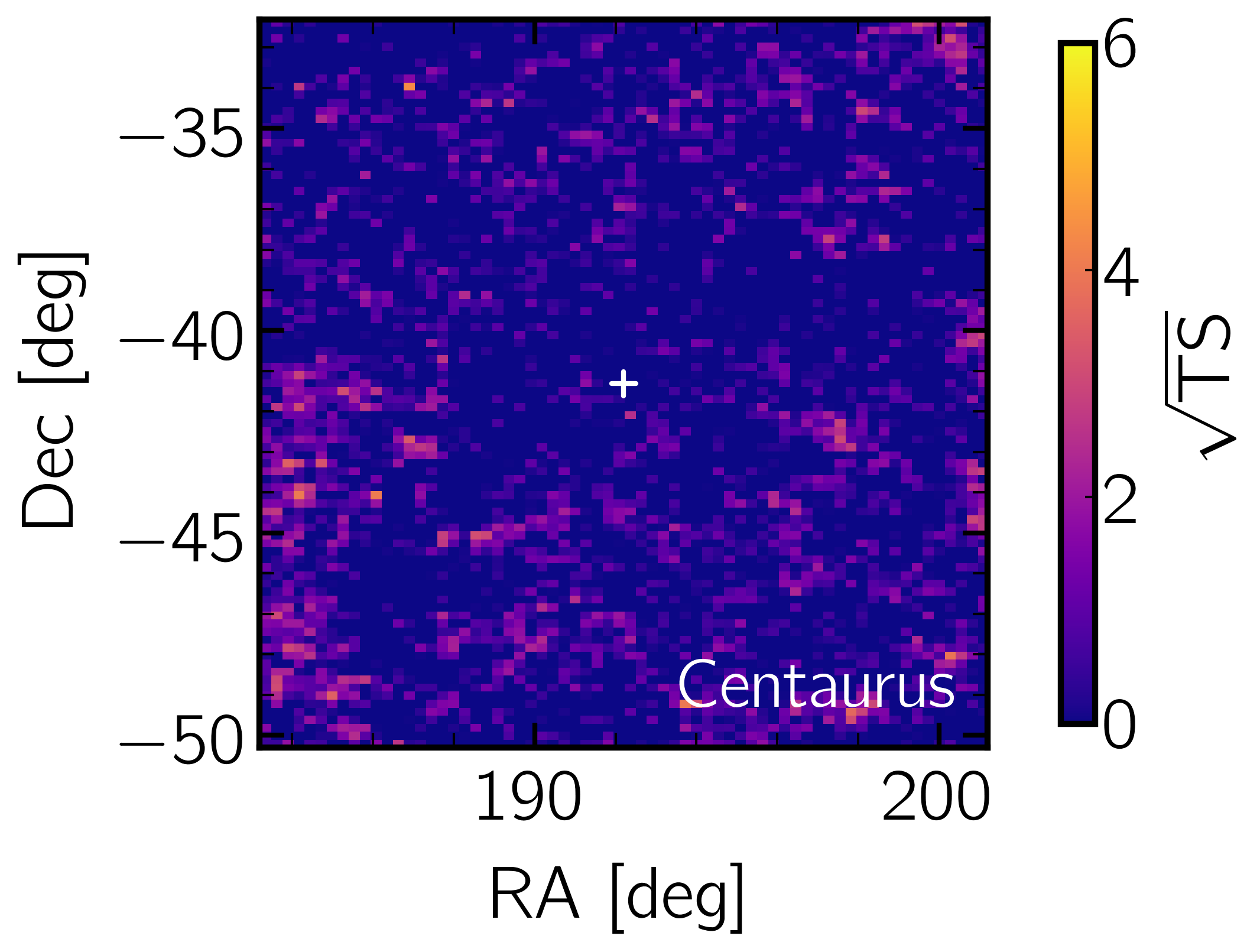} &
\includegraphics[width=0.32\textwidth]{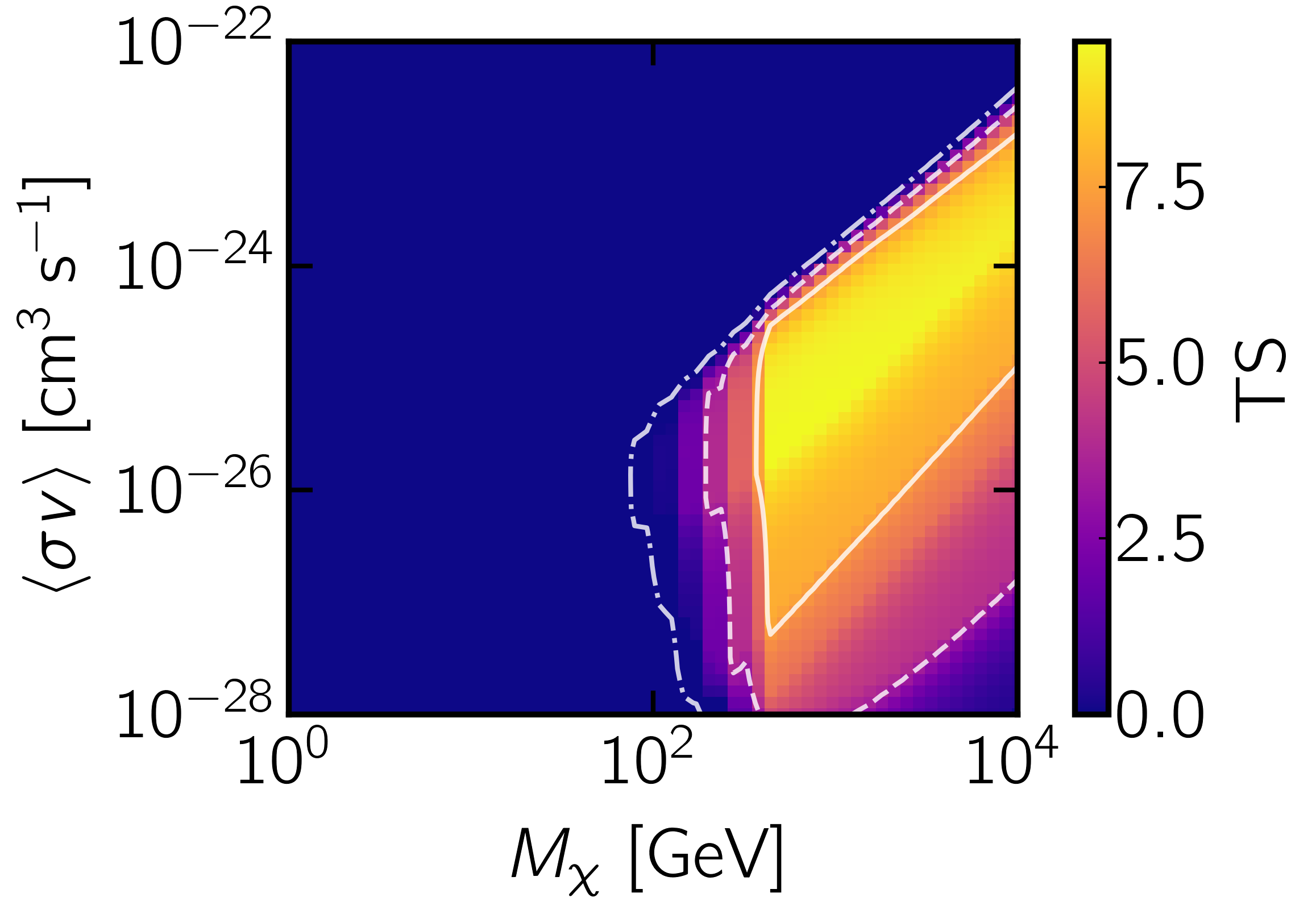} &
\includegraphics[width=0.32\textwidth]{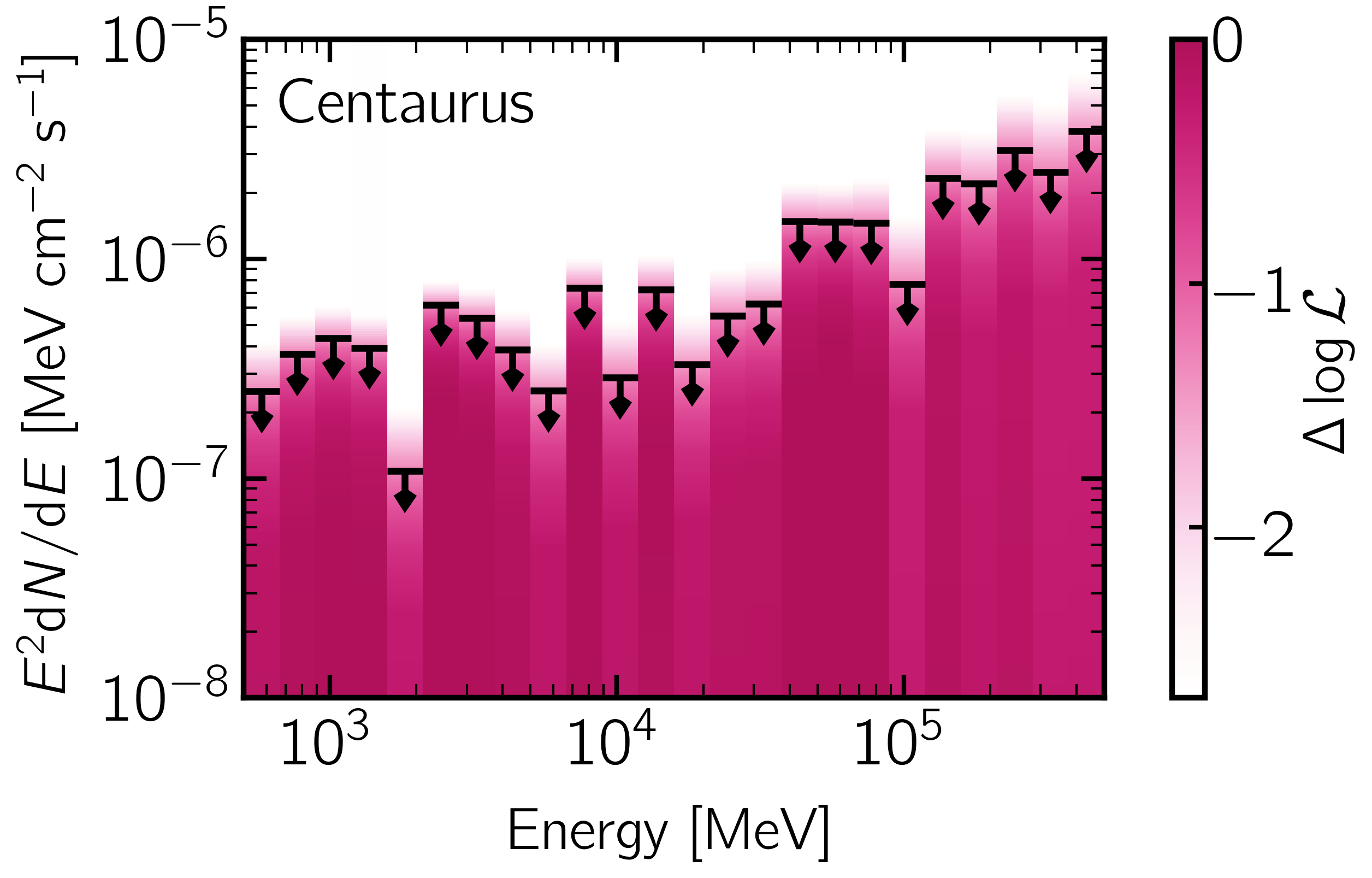} \\[0.3cm]
\end{tabular}
\caption{}
\label{fig:lat_results2}
\end{figure*}
\renewcommand\thefigure{\arabic{figure}}
    \setcounter{figure}{7}

\subsection{\textit{Fermi}-LAT results}
Figure~\ref{fig:lat_results} presents comprehensive \textit{Fermi}-LAT analysis results for all seven galaxy clusters. All results shown here are computed using the fixed $J$-factor values listed in Table~\ref{tab:clusters} (this is in contrast to our main results shown in Fig.~\ref{fig:ul_clusters} which account for the $\mDM$-dependent $J$-factors shown in Fig.~\ref{fig:cluster-J}). We find no statistically significant $\gamma$-ray sources at cluster locations reported in Table~\ref{tab:clusters}. The way to interpret Fig.~\ref{fig:lat_results} is as follows:

\begin{itemize}
\item \textbf{Left column (TS Maps):} Test statistic ($\sqrt{\text{TS}}$) maps assess $\gamma$-ray emission significance across each cluster's RoI. These maps assume a power-law spectral model with fixed photon index ($\Gamma=2.0$) but free normalization, serving as a proxy for the strength of $\gamma$-ray emission at each spatial location covered by the DM annihilation map derived in Sec.~\ref{sec:cusps}. The white cross marks each cluster's dynamical center, and the color scale indicates the square root of the TS, which approximates the detection significance in units of $\sigma$. Values above $\sqrt{\text{TS}} \gtrsim 3$ would indicate potential $\gamma$-ray sources warranting further investigation.

\item \textbf{Middle column (DM Constraints):} DM parameter space constraints show the TS as a function of dark matter mass ($m_{\text{DM}}$) and velocity-averaged annihilation cross-section ($\langle\sigma v\rangle$) in Eq.~\ref{eq:tsDM}, for the $b\bar{b}$ annihilation channel, computed using each cluster's specific $J$-factor from Table~\ref{tab:clusters}. Here, TS follows a $\chi^2$ distribution with 2 degrees of freedom (DM mass and velocity-weightet annihilation cross-section), and the contours indicate 68\%, 95\%, and 99.7\% CL. White solid contours delineate the strongest exclusion regions for the assumed $J$-factor values.

\item \textbf{Right column (SEDs):} Spectral energy distributions display the differential $\gamma$-ray flux ($E^2 \text{d}N/\text{d}E$) as a function of energy from 500 MeV to 500 GeV. Black arrows indicate 95\% confidence ULs for energy bins where no significant emission is detected (TS $\lesssim 9$), while error bars show flux measurements with uncertainties for bins with modest evidence (TS $\gtrsim 9$). The pink colored likelihood bands represent $\Delta \log \mathcal{L}$ profiles across energy, where TS (middle column) $= -2\Delta \log \mathcal{L}$, summed over all energy bins. For clusters showing elevated TS values, the blue curves indicate the best-fit DM annihilation spectra for the assumed $J$-factor and the $b\bar{b}$ annihilation channel.
\end{itemize}

\subsection{Notes on individual clusters}

\paragraph*{\textbf{Virgo Cluster.}}
The Virgo region contains multiple $\gamma$-ray components. M87 (4FGL~J1230.2+1223) is an established point source with average flux  $F_\gamma \sim 1.6470 \times 10^{-9}$~ph~cm$^{-2}$~s$^{-1}$ ($1~\text{GeV}< E <100$~GeV) and photon index $\Gamma = 2.06 \pm 0.04$ \cite{Fermi-LAT:2022byn}. It is also a flaring source: the EHT Collaboration detected the first $\gamma$-ray flare in over a decade during their 2018 EHT campaign \cite{EventHorizonTelescope:2024uoo}. Additionally, Ackermann \textit{et al}. reported a statistically significant extended emission with uniformly emitting disk profile (radius = 3$^{\circ}$), though its significance strongly depends on the adopted interstellar emission model and ``is most likely an artifact of our incomplete description of the IEM in this region.'' ~\cite{Fermi-LAT:2015xij}

Our analysis shows no significant localized $\gamma$-ray excess across the entire RoI. The TS map reveals some scattered regions with $\sqrt{\text{TS}}\sim4$--$6$, which could indicate either statistical fluctuations or diffuse $\gamma$-ray emission, but we find no coherent extended structure correlated with our DM template. The DM parameter space exhibits a maximum TS $\sim 15$ around $m_{\text{DM}} \sim 2.5$~TeV (corresponding to $\sim$3.9$\sigma$ for 2 d.o.f.). The SED shows consistent upper limits across all energy bins. While these features can be suggestive of potential $\gamma$-ray activity, they do not result in a robust detection. Thus, the Virgo cluster, unsurprisingly, provides the strongest constraints among our targets, primarily due its proximity and large annihilation $J$-factor (Fig.~\ref{fig:ul_clusters}). 

\paragraph*{\textbf{NGC~4636.}} NGC~4636 is a giant elliptical galaxy within the Virgo supercluster. Our analysis finds no significant $\gamma$-ray excess. A TS $\sim$ 30 preference around $m_{\text{DM}} \sim$ 441~GeV for the $b\bar{b}$ annihilation channel appears in our SED analysis. This represents the highest-significance preference for a DM annihilation spectral model in our cluster sample, though we note that the signal remains consistent with astrophysical origins (especially considering the crowded environment of NGC~4636). 

\paragraph*{\textbf{M49.}} M49 is a giant elliptical galaxy located within the Virgo supercluster. Our analysis reveals no significant $\gamma$-ray excess associated with its expected DM template. The maximum TS associated with the DM parameter space is $\sim$ 6. The SED analysis yields robust upper limits across all energy bins.

\paragraph*{\textbf{Fornax (NGC~1399).}} The Fornax cluster, centered on the elliptical galaxy NGC~1399, shows no excess $\gamma$-ray emission above background expectation. When fitting the DM annihilation spectrum to the observed flux limits, the data exhibit a modest statistical preference for DM model, with a peak TS $\sim12$ (corresponding to $\sim 3.5\sigma$). The SED analysis identifies a best-fit DM mass of $\sim$ 962 GeV for annihilation into the $b\bar{b}$ channel. 

\paragraph*{\textbf{Hydra Cluster.}} Our analysis does not detect any significant $\gamma$-ray excess, with DM spectral fitting yielding a maximum TS value of $\sim$7 . The SED analysis confirms that flux upper limits are consistent with background expectations across all energies. 

\paragraph*{\textbf{Coma Cluster.}} The Coma cluster has confirmed $\gamma$-ray emission from multiple independent analysis \cite{Xi:2017uzz, Adam:2021dsu, Baghmanyan:2021jwg}. Adam \textit{et al.} confirm detection with TS $\sim$ 27 using a binned likelihood approach, with $\gamma$-ray morphology elongated in the northeast--southwest direction coincident with a radio halo \cite{Adam:2021dsu}. Additionally, a point source 4FGL~J1256.9+2736 about 0.8$^{\circ}$ from the Coma's dynamical center has a TS $\sim$ 17 in the 4FGL catalog~\cite{Fermi-LAT:2022byn}, which could either be astrophysical contamination or correspond to the peak of diffuse cluster emission.

We find no statistically significant $\gamma$-ray excess coincident with Coma's region that can be attributed to cuspy DM. When fitting the DM annihilation spectra to the observed upper limits on flux, the analysis yields a maximum TS value exceeding 25, corresponding to approximately 5$\sigma$ significance for the DM model preference. The SED analysis identifies a best-fit DM mass of about 441~GeV for annihilation into the $b\bar{b}$ channel. While this apparent alignment with a theoretical DM annihilation spectrum is notable, the emission could also originate from known astrophysical sources or processes associated with the cluster's radio halo. Hence, systematic uncertainties and potential astrophysical contributions must be thoroughly considered before attributing this excess TS conclusively to DM annihilation. 

\paragraph*{\textbf{Centaurus Cluster.}} Centaurus hosts extended $\gamma$-ray emission from giant lobes of Centaurus, with emission extending beyond radio image to energies up to 30 GeV~\cite{Sun:2016ibh}. We find no statistically significant excess in our analysis, with a DM model maximum TS value of $\sim 8$. Similar to other clusters in our sample, the SED analysis returns flux upper limits comparable with background fluctuations across the energy range considered.

\subsection{Concluding remarks}

Our comprehensive analysis of seven galaxy clusters reveals no significant $\gamma$-ray detections above background expectations. However, when fitting DM annihilation spectra to the observed flux limits, several clusters show modest statistical preferences for DM models over background-only scenarios. The closest and best characterized system---the Virgo cluster---shows a maximum TS $\sim$15 around $m_{\text{DM}}\sim 2.5$~TeV. The Fornax cluster exhibits TS $\sim$ 12 ($\sim$3.5$\sigma$) with a best-fit mass of $\sim$ 962~GeV for $b\bar{b}$ annihilation, while both Coma and NGC~4636 present the highest TS values with TS $>$ 25 ($\sim$5$\sigma$ significance) and best-fit mass of $\sim$ 441~GeV.

These TS values are not necessarily consistent with a DM interpretation when considered collectively. Instead, they likely originate from known astrophysical processes \cite{Fermi-LAT:2015xij, Fermi-LAT:2022byn, EventHorizonTelescope:2024uoo, Xi:2017uzz, Adam:2021dsu, Baghmanyan:2021jwg, Sun:2016ibh}: Coma's signal can correlate with its confirmed radio halo and diffuse emission. Notably, NGC~4636, Virgo, and M49 are all members of the Virgo supercluster, suggesting that any large-scale astrophysical processes affecting one system could extend to the others and provide a common origin for observed $\gamma$-ray activity. Critically, all putative DM signals from other clusters are excluded by the robust upper limits derived from Virgo---our best characterized system with the largest $J$-factor and proximity.

\bibliography{main}

\end{document}